\documentclass[twocolumn,pre,showpacs,superscriptaddress]{revtex4-1}%Gian's line
\usepackage{graphicx,color}
\usepackage{amsmath,amssymb,latexsym,amsthm}
\usepackage{mathtools}
\usepackage{tikz}
\usetikzlibrary{chains,shapes,fit,calc,arrows}
\usepackage{pgfplots, pgfplotstable}
\usepgfplotslibrary{groupplots}

\usepackage{color}

\definecolor{webgreen}{rgb}{0,.5,0}
\definecolor{webbrown}{rgb}{.6,0,0}
\definecolor{grigio}{rgb}{.85,.85,.85} 
\definecolor{RoyalBlue}{rgb}{0.0, 0.14, 0.4}
\definecolor{skyblue1}{rgb}{0.45,0.62,0.81}
\definecolor{skyblue2}{rgb}{0.2,0.39,0.64}
\definecolor{skyblue3}{rgb}{0.13,0.29,0.53}
\definecolor{scarlet1}{rgb}{0.93,0.16,0.16}
\definecolor{scarlet2}{rgb}{0.8,0,0}
\definecolor{scarlet3}{rgb}{0.64,0,0}

\begin{document}

\title{The Thermo-Kinetic Relations}
\newcommand\unipd{\affiliation{Department of Physics and Astronomy, University of Padova, Via Marzolo 8, I-35131 Padova, Italy.}}
\newcommand\unilou{\affiliation{Institute of Information and Communication Technologies, Electronics and Applied Mathematics Universit\'{e} catholique de Louvain, Louvain-La-Neuve, Belgium
}}
\author{Jean-Charles Delvenne}
\unilou
\author{Gianmaria Falasco}
\unipd

%\footnote{J-C D acknowledges funding from the Project INTER/FNRS/20/15074473 ‘TheCirco’ on Thermodynamics of Circuits for Computation, funded by the F.R.S.-FNRS (Belgium) and FNR (Luxembourg).}

\begin{abstract}

Thermo-Kinetic relations bound thermodynamic quantities such as entropy production with statistics of dynamical observables. We introduce a Thermo-Kinetic Relation to bound the entropy production or the non-adiabatic (Hatano-Sasa, excess) entropy production for overdamped Markov jump processes, possibly with time-varying rates and non stationary distributions.
For stationary cases, this bound is akin to a Thermodynamic Uncertainty Relation, only involving absolute fluctuations rather than the mean square, thereby offering a better lower bound far from equilibrium. For non-stationary cases, this bound generalises Classical Speed Limits, where the kinetic term is not necessarily the activity (number of jumps) but any %symmetric or antisymmetric 
trajectory observable of interest. 
As a consequence, in the task of driving a system from a given probability distribution to another, we find a trade-off between non-adiabatic entropy production and housekeeping entropy production: the latter can be increased in order to decrease the former, although to a limited extent. We also find constraints specific to constant-rate Markov processes. 
We illustrate our Thermo-Kinetic Relations on simple examples from biophysics and computing devices. 
\end{abstract}


\maketitle

Suppose the state of a stochastic system is driven from a probability distribution to another via a Markovian protocol, i.e. a time-varing sequence of transition rates. 
 Classical Speed Limits establish bounds between the time taken by the protocol, the statistical distance between initial and final state, and entropy production---or, preferably, the Hatano-Sasa or non-adiabatic (NA) \cite{hat01, esposito2010three} entropy production as it leads to tighter bounds. The first Classical Speed Limit for discrete state spaces  
was established in  \cite{shi18}, later improved or generalised by \cite{vo20,vanvu21}. However, it was not known how tight these bounds were. Recently, a version involving entropy production was proved tight \cite{dechant2022minimum}, in that it is achievable by a specific protocol. In this note, we achieve a Classical Speed Limit that is close to tightness, in that it cannot be improved by more than 18 percent. We show in particular that non-conservative forces are useful in order to lower non-adiabatic entropy production to its lowest possible level, in contrast with minimum entropy production protocols.  We also prove that for constant protocols, in some cases the entropy production has an incompressible lower bound even regardless of the time or activity of the transformation. 

In fact our bounds are more general than the above suggests, as they relate entropy production, not just to activity, but to any observable, symmetric or antisymmetric under time-reversal, along the trajectories. They also relate  non-adiabatic entropy production to any trajectory observable with some flow preservation property. 
  
These general Thermo-Kinetic Relations also offer novel bounds on the entropy production for stationary (or periodic) non-equilibrium Markov processes, where they resemble Thermodynamic Uncertainty Relations. The latter were introduced as an inequality between entropy production and mean-to-standard-deviation of a time-antisymmetric observable  \cite{bar15a, GingHorEng16, pol16, hor17, pro17, gin17fp, dechant18current, potts2019thermodynamic, hasegawa19, van20, liu20, falasco2020unifying}. In contrast, the present bounds involve the absolute value of a fluctuating observable, rather than the mean square. Numerical and theoretical comparisons indicate that the bound on entropy production is tighter than the well-known Thermodynamic Uncertainty Relation at least far from equilibrium.    
  
The article is organised as follows. In Section \ref{sec:1} we reformulate elementary concepts related to entropy production for arbitrary  time-varying or constant Markovian protocols. We prove in Section \ref{sec:2} our main result, a Thermo-Kinetic Relation for entropy production, for any observable along the trajectories, first for constant-rate Markov chains, then, via integration over infinitesimal intervals, for time-varying Markov chains. In Section \ref{sec:3} we show the connection with optimal transport theory, recovering the speed limit for general Kantorovich costs. The case of NA entropy production is explored in Section \ref{sec:4}. In the last section, we discuss consequences and applications.

\section{Useful concepts of probability theory and stochastic thermodynamics}

\label{sec:1}

 \subsection{Kullback-Leibler divergence and total variation distance}
 
 Before any physics, we start with mathematical definitions and  facts on divergences. Given an arbitrary probability space $\Omega$ endowed with two probability distribution $p$ and $q$, the Kullback-Leibler of $p$ relative to $q$ is a nonnegative, possibly infinite, real number defined as
\begin{align}
	D(p\|q)= \sum_{\omega} p(\omega) \ln \frac{p(\omega)}{q(\omega)}
\end{align} 

We adopt the sum notation for simplicity even though the definition makes sense for continuous and discrete distributions alike. In the former case, the sum is to be replaced with an integral and the ratio of probabilities with a probability density function.

The following well-known identity (Chain Rule for Kullback-Leibler divergence) will be repeatedly useful. Consider  two probability distributions $p$ and $q$ over an arbitrary space $\Omega$ and an observable $X:\Omega \to \mathcal{X}$, taking values in an arbitrary set $\mathcal{X}$. Then we have the following decomposition into two nonnegative terms:
\begin{align}
	D(p\|q)= D(p_X\|q_X)+D(p\|q|X) .
\end{align}
Here $p_X$  is the probability distribution on the observable $X$, derived from probability distribution $p$ on $\Omega$,   that is $p_X(x)= \sum_{\omega:  X(\omega)=x}  p(\omega)$. The conditional Kullback-Leibler divergence is defined as 
\begin{align}
	D(p\|q|X) &= \sum_x p_X(x) D(p\|q|X=x) \\
	&= \sum_x p_X(x) \sum_{\omega: X(\omega)=x} \frac{p(\omega)}{p_X(x)} \ln  \frac{p(\omega)/p_X(x)}{q(\omega)/q_X(x)}.\nonumber
\end{align}

We introduce a simple bound for Kullback-Leibler divergences. Consider two arbitrary probability distributions $p$ and $q$ on the space $\Omega$, and  a nonnegative function $A:\Omega \to \mathbb{R}^+$ such that $p_A=q_A$ (in other words, the event $A=a$ has the same probability under $p$ and under $q$, for all $a$). Then we can define  $\varphi=A/\langle A \rangle$ as a probability density function for both $p$ and $q$, meaning that both $\varphi p$ and $\varphi q$ are valid probability distributions. Then we can write: 

\begin{align}\label{eq:Dpqphi}
	D(\varphi p\|\varphi  q) \leq \varphi_{\max}  	D( p\|  q), 
\end{align}
where $\varphi_{\max}=A_{\max}/\langle A \rangle$ is the maximum (or supremum) of $\varphi$ over $\Omega$. In other words:

\begin{align}\label{eq:DpqA}
	D( p\|  q) \geq \frac{\langle A \rangle}{A_{\max}}D\left(\frac{Ap}{\langle A \rangle}\|\frac{Aq}{\langle A \rangle}\right).
\end{align}

Equality holds when $A$ takes only two values $0$ and $A_{\max}$, with $p=q$ for all $\omega$ such that $A(\omega)=0$.
The proof goes as follows:
\begin{align}
	D(\varphi p\|\varphi  q) &= D(\varphi p\|\varphi  q | \varphi)\\ 
							&= \sum_{z} zp(\varphi=z)  D(\varphi p\|\varphi  q | \varphi=z) \label{eq:sumDphi}\\
							&\leq \varphi_{\max} \sum_{z} p(\varphi=z)  D(\varphi p\|\varphi  q | \varphi=z)\\
							&= \varphi_{\max} \sum_{z} p(\varphi=z) \!\!\! \sum_{\omega:\varphi(\omega)=z} \!\!\! \frac{z p(\omega)}{z p(\varphi=z)} \ln \frac{zp(\omega)}{zq(\omega)}\\
							& = \varphi_{\max}   D( p\|  q | \varphi)\\
							& = \varphi_{\max}   D( p\|  q ).
\end{align}

Note that in case of an infinite space $\Omega$, we may have $\varphi$ taking arbitrarily high values with vanishingly small probabilities, in which case we can replace $\varphi_{\max}$ with an `effective' maximum, see Appendix.

Another notion of divergence between two probability distributions is the total variation distance:

\begin{align}
	d_{TV}(p,q)=\frac{1}{2}\sum_{\omega} |p(\omega)-q(\omega)| \leq 1.
\end{align}
Unlike the Kullback-Leibler divergence, this is a proper distance---in particular, it is symmetric and respects the triangle inequality. It is related to the Kullback-Leibler distance in the following way:
We can find a convex function $h$ with $h(0)=0$ such that for every distribution $p,q$  on some space we have:
\begin{align} \label{eq:DpqhdTV}
	D(p \|q) \geq  h(d_{TV}(p,q)).
\end{align}
Pinsker's inequality states that $x \mapsto h(x)= 2x^2$ is such a function. Although well known, Pinsker's inequality is far from optimal. Vajda's bound \cite{vajda1970note} $h(x)=\ln\frac{1+x}{1-x} - 2\frac{x}{1+x}$ is much tighter for $d_{TV}(p,q) \approx 1$ where Vajda's inequality correctly predicts an unbounded Kullback-Leibler divergence, unlike Pinsker's inequality. Even better is Gilardoni's bound \cite{gilardoni2008improvement}, for
\begin{align}\label{eq:gilard}
	h(x)&=\ln\frac{1}{1-x} - (1-x)\ln(1+x)\\
	&=\ln\frac{1+x}{1-x} - (2-x)\ln(1+x).
\end{align}
Note that Pinsker's inequality remains very slightly better than Gilardoni's for small values $d_{TV}(p,q) \approx 0$.

There is an optimal such convex function $h^*$, beating in particular Pinsker, Vajda and Gilardoni's bounds. It is optimal in that for each $d \in [0,1]$, there exist  distributions  $p,q$ such that  $d_{TV}(p,q)=d$ and $D(p\|q)=h^*(d)$. Interestingly, these optimal distributions can always be chosen on two states only. Despite this, $h^*$ has no known explicit analytic expression, although it can easily be computed numerically through an implicit formula \cite{fedotov2003refinements}. 

\begin{figure}
\center
\begin{tikzpicture}
\begin{axis}[xlabel=$x$, ylabel={$h(x)$},
    xmin = 0, xmax = 1,
    ymin = 0, ymax = 3,legend style={at={(0.05,0.75)},anchor=west}]
    \addplot[samples = 200, smooth,thick,teal, dotted] {2*x*x};
     \addlegendentry{Pinsker};
%    \addplot[samples = 200, smooth,thick,blue, dashed]  {ln((1 + x)/(1 - x))-2*x/(1+x)};
%     \addlegendentry{Vajda};
     \addplot[samples = 200, smooth,thick,red, dashdotted]  {ln(1/(1 - x)) - (1 - x)*ln(1 + x)};
       \addlegendentry{Gilardoni};
     \addplot[mark=none,thick,dashed,blue] coordinates {
     (0.,0)(0.0005,5.*10^-7)(0.001,2.*10^-6)(0.0015,4.5*10^-6)(0.002,8.00001*10^-6)(0.0025,0.0000125)(0.003,0.000018)(0.0035,0.0000245001)(0.004,0.0000320001)(0.0045,0.0000405002)(0.005,0.0000500003)(0.0055,0.0000605004)(0.006,0.0000720006)(0.0065,0.0000845008)(0.007,0.0000980011)(0.0075,0.000112501)(0.008,0.000128002)(0.0085,0.000144502)(0.009,0.000162003)(0.0095,0.000180504)(0.01,0.000200004)(0.0105,0.000220505)(0.011,0.000242007)(0.0115,0.000264508)(0.012,0.000288009)(0.0125,0.000312511)(0.013,0.000338013)(0.0135,0.000364515)(0.014,0.000392017)(0.0145,0.00042052)(0.015,0.000450023)(0.0155,0.000480526)(0.016,0.000512029)(0.0165,0.000544533)(0.017,0.000578037)(0.0175,0.000612542)(0.018,0.000648047)(0.0185,0.000684552)(0.019,0.000722058)(0.0195,0.000760564)(0.02,0.000800071)(0.0205,0.000840579)(0.021,0.000882086)(0.0215,0.000924595)(0.022,0.000968104)(0.0225,0.00101261)(0.023,0.00105812)(0.0235,0.00110464)(0.024,0.00115215)(0.0245,0.00120066)(0.025,0.00125017)(0.0255,0.00130069)(0.026,0.0013522)(0.0265,0.00140472)(0.027,0.00145824)(0.0275,0.00151275)(0.028,0.00156827)(0.0285,0.00162479)(0.029,0.00168231)(0.0295,0.00174084)(0.03,0.00180036)(0.0305,0.00186088)(0.031,0.00192241)(0.0315,0.00198494)(0.032,0.00204847)(0.0325,0.002113)(0.033,0.00217853)(0.0335,0.00224506)(0.034,0.00231259)(0.0345,0.00238113)(0.035,0.00245067)(0.0355,0.00252121)(0.036,0.00259275)(0.0365,0.00266529)(0.037,0.00273883)(0.0375,0.00281338)(0.038,0.00288893)(0.0385,0.00296548)(0.039,0.00304303)(0.0395,0.00312158)(0.04,0.00320114)(0.0405,0.0032817)(0.041,0.00336326)(0.0415,0.00344582)(0.042,0.00352938)(0.0425,0.00361395)(0.043,0.00369952)(0.0435,0.00378609)(0.044,0.00387367)(0.0445,0.00396224)(0.045,0.00405182)(0.0455,0.00414241)(0.046,0.00423399)(0.0465,0.00432658)(0.047,0.00442017)(0.0475,0.00451477)(0.048,0.00461036)(0.0485,0.00470696)(0.049,0.00480457)(0.0495,0.00490317)(0.05,0.00500278)(0.0505,0.00510339)(0.051,0.00520501)(0.0515,0.00530763)(0.052,0.00541125)(0.0525,0.00551588)(0.053,0.00562151)(0.0535,0.00572815)(0.054,0.00583579)(0.0545,0.00594443)(0.055,0.00605407)(0.0555,0.00616472)(0.056,0.00627638)(0.0565,0.00638904)(0.057,0.0065027)(0.0575,0.00661737)(0.058,0.00673304)(0.0585,0.00684971)(0.059,0.0069674)(0.0595,0.00708608)(0.06,0.00720577)(0.0605,0.00732647)(0.061,0.00744817)(0.0615,0.00757087)(0.062,0.00769458)(0.0625,0.0078193)(0.063,0.00794502)(0.0635,0.00807174)(0.064,0.00819947)(0.0645,0.00832821)(0.065,0.00845795)(0.0655,0.0085887)(0.066,0.00872045)(0.0665,0.00885321)(0.067,0.00898698)(0.0675,0.00912175)(0.068,0.00925753)(0.0685,0.00939431)(0.069,0.0095321)(0.0695,0.0096709)(0.07,0.0098107)(0.0705,0.00995151)(0.071,0.0100933)(0.0715,0.0102361)(0.072,0.01038)(0.0725,0.0105248)(0.073,0.0106707)(0.0735,0.0108175)(0.074,0.0109654)(0.0745,0.0111142)(0.075,0.0112641)(0.0755,0.011415)(0.076,0.0115669)(0.0765,0.0117198)(0.077,0.0118737)(0.0775,0.0120286)(0.078,0.0121845)(0.0785,0.0123414)(0.079,0.0124994)(0.0795,0.0126583)(0.08,0.0128183)(0.0805,0.0129792)(0.081,0.0131412)(0.0815,0.0133042)(0.082,0.0134682)(0.0825,0.0136332)(0.083,0.0137992)(0.0835,0.0139662)(0.084,0.0141342)(0.0845,0.0143032)(0.085,0.0144733)(0.0855,0.0146443)(0.086,0.0148164)(0.0865,0.0149895)(0.087,0.0151636)(0.0875,0.0153387)(0.088,0.0155148)(0.0885,0.0156919)(0.089,0.01587)(0.0895,0.0160491)(0.09,0.0162293)(0.0905,0.0164104)(0.091,0.0165926)(0.0915,0.0167758)(0.092,0.01696)(0.0925,0.0171452)(0.093,0.0173314)(0.0935,0.0175186)(0.094,0.0177069)(0.0945,0.0178961)(0.095,0.0180864)(0.0955,0.0182776)(0.096,0.0184699)(0.0965,0.0186632)(0.097,0.0188575)(0.0975,0.0190529)(0.098,0.0192492)(0.0985,0.0194466)(0.099,0.0196449)(0.0995,0.0198443)(0.1,0.0200447)(0.1005,0.0202461)(0.101,0.0204485)(0.1015,0.0206519)(0.102,0.0208564)(0.1025,0.0210618)(0.103,0.0212683)(0.1035,0.0214758)(0.104,0.0216843)(0.1045,0.0218938)(0.105,0.0221043)(0.1055,0.0223159)(0.106,0.0225284)(0.1065,0.022742)(0.107,0.0229566)(0.1075,0.0231722)(0.108,0.0233888)(0.1085,0.0236065)(0.109,0.0238251)(0.1095,0.0240448)(0.11,0.0242655)(0.1105,0.0244872)(0.111,0.0247099)(0.1115,0.0249337)(0.112,0.0251584)(0.1125,0.0253842)(0.113,0.025611)(0.1135,0.0258388)(0.114,0.0260676)(0.1145,0.0262974)(0.115,0.0265283)(0.1155,0.0267602)(0.116,0.0269931)(0.1165,0.027227)(0.117,0.0274619)(0.1175,0.0276978)(0.118,0.0279348)(0.1185,0.0281728)(0.119,0.0284118)(0.1195,0.0286518)(0.12,0.0288929)(0.1205,0.0291349)(0.121,0.029378)(0.1215,0.0296221)(0.122,0.0298672)(0.1225,0.0301134)(0.123,0.0303606)(0.1235,0.0306087)(0.124,0.0308579)(0.1245,0.0311082)(0.125,0.0313594)(0.1255,0.0316117)(0.126,0.031865)(0.1265,0.0321193)(0.127,0.0323746)(0.1275,0.032631)(0.128,0.0328884)(0.1285,0.0331468)(0.129,0.0334062)(0.1295,0.0336666)(0.13,0.0339281)(0.1305,0.0341906)(0.131,0.0344541)(0.1315,0.0347186)(0.132,0.0349842)(0.1325,0.0352508)(0.133,0.0355184)(0.1335,0.035787)(0.134,0.0360567)(0.1345,0.0363274)(0.135,0.0365991)(0.1355,0.0368718)(0.136,0.0371456)(0.1365,0.0374203)(0.137,0.0376962)(0.1375,0.037973)(0.138,0.0382508)(0.1385,0.0385297)(0.139,0.0388096)(0.1395,0.0390906)(0.14,0.0393725)(0.1405,0.0396555)(0.141,0.0399396)(0.1415,0.0402246)(0.142,0.0405107)(0.1425,0.0407978)(0.143,0.0410859)(0.1435,0.0413751)(0.144,0.0416652)(0.1445,0.0419565)(0.145,0.0422487)(0.1455,0.042542)(0.146,0.0428363)(0.1465,0.0431316)(0.147,0.043428)(0.1475,0.0437253)(0.148,0.0440238)(0.1485,0.0443232)(0.149,0.0446237)(0.1495,0.0449252)(0.15,0.0452277)(0.1505,0.0455313)(0.151,0.0458359)(0.1515,0.0461415)(0.152,0.0464482)(0.1525,0.0467559)(0.153,0.0470646)(0.1535,0.0473744)(0.154,0.0476852)(0.1545,0.047997)(0.155,0.0483099)(0.1555,0.0486238)(0.156,0.0489387)(0.1565,0.0492547)(0.157,0.0495716)(0.1575,0.0498897)(0.158,0.0502087)(0.1585,0.0505288)(0.159,0.05085)(0.1595,0.0511721)(0.16,0.0514953)(0.1605,0.0518196)(0.161,0.0521448)(0.1615,0.0524711)(0.162,0.0527985)(0.1625,0.0531269)(0.163,0.0534563)(0.1635,0.0537867)(0.164,0.0541182)(0.1645,0.0544507)(0.165,0.0547843)(0.1655,0.0551189)(0.166,0.0554545)(0.1665,0.0557912)(0.167,0.0561289)(0.1675,0.0564677)(0.168,0.0568075)(0.1685,0.0571483)(0.169,0.0574902)(0.1695,0.0578331)(0.17,0.058177)(0.1705,0.058522)(0.171,0.0588681)(0.1715,0.0592151)(0.172,0.0595633)(0.1725,0.0599124)(0.173,0.0602626)(0.1735,0.0606138)(0.174,0.0609661)(0.1745,0.0613194)(0.175,0.0616738)(0.1755,0.0620292)(0.176,0.0623857)(0.1765,0.0627431)(0.177,0.0631017)(0.1775,0.0634613)(0.178,0.0638219)(0.1785,0.0641835)(0.179,0.0645463)(0.1795,0.06491)(0.18,0.0652748)(0.1805,0.0656407)(0.181,0.0660075)(0.1815,0.0663755)(0.182,0.0667445)(0.1825,0.0671145)(0.183,0.0674856)(0.1835,0.0678577)(0.184,0.0682309)(0.1845,0.0686051)(0.185,0.0689803)(0.1855,0.0693566)(0.186,0.069734)(0.1865,0.0701124)(0.187,0.0704919)(0.1875,0.0708724)(0.188,0.0712539)(0.1885,0.0716365)(0.189,0.0720202)(0.1895,0.0724049)(0.19,0.0727906)(0.1905,0.0731775)(0.191,0.0735653)(0.1915,0.0739542)(0.192,0.0743442)(0.1925,0.0747352)(0.193,0.0751272)(0.1935,0.0755204)(0.194,0.0759145)(0.1945,0.0763097)(0.195,0.076706)(0.1955,0.0771033)(0.196,0.0775017)(0.1965,0.0779012)(0.197,0.0783016)(0.1975,0.0787032)(0.198,0.0791058)(0.1985,0.0795094)(0.199,0.0799141)(0.1995,0.0803199)(0.2,0.0807267)(0.2005,0.0811346)(0.201,0.0815435)(0.2015,0.0819535)(0.202,0.0823646)(0.2025,0.0827767)(0.203,0.0831898)(0.2035,0.0836041)(0.204,0.0840193)(0.2045,0.0844357)(0.205,0.0848531)(0.2055,0.0852715)(0.206,0.085691)(0.2065,0.0861116)(0.207,0.0865332)(0.2075,0.0869559)(0.208,0.0873797)(0.2085,0.0878045)(0.209,0.0882304)(0.2095,0.0886573)(0.21,0.0890853)(0.2105,0.0895144)(0.211,0.0899445)(0.2115,0.0903757)(0.212,0.090808)(0.2125,0.0912413)(0.213,0.0916757)(0.2135,0.0921111)(0.214,0.0925476)(0.2145,0.0929852)(0.215,0.0934239)(0.2155,0.0938636)(0.216,0.0943044)(0.2165,0.0947462)(0.217,0.0951891)(0.2175,0.0956331)(0.218,0.0960781)(0.2185,0.0965242)(0.219,0.0969714)(0.2195,0.0974196)(0.22,0.097869)(0.2205,0.0983193)(0.221,0.0987708)(0.2215,0.0992233)(0.222,0.0996769)(0.2225,0.100132)(0.223,0.100587)(0.2235,0.101044)(0.224,0.101502)(0.2245,0.101961)(0.225,0.102421)(0.2255,0.102882)(0.226,0.103344)(0.2265,0.103807)(0.227,0.104272)(0.2275,0.104737)(0.228,0.105204)(0.2285,0.105671)(0.229,0.10614)(0.2295,0.106609)(0.23,0.10708)(0.2305,0.107552)(0.231,0.108025)(0.2315,0.108499)(0.232,0.108974)(0.2325,0.10945)(0.233,0.109927)(0.2335,0.110406)(0.234,0.110885)(0.2345,0.111365)(0.235,0.111847)(0.2355,0.11233)(0.236,0.112813)(0.2365,0.113298)(0.237,0.113784)(0.2375,0.114271)(0.238,0.114759)(0.2385,0.115248)(0.239,0.115738)(0.2395,0.116229)(0.24,0.116722)(0.2405,0.117215)(0.241,0.11771)(0.2415,0.118205)(0.242,0.118702)(0.2425,0.1192)(0.243,0.119699)(0.2435,0.120199)(0.244,0.1207)(0.2445,0.121202)(0.245,0.121705)(0.2455,0.122209)(0.246,0.122715)(0.2465,0.123221)(0.247,0.123729)(0.2475,0.124237)(0.248,0.124747)(0.2485,0.125258)(0.249,0.12577)(0.2495,0.126283)(0.25,0.126797)(0.2505,0.127312)(0.251,0.127828)(0.2515,0.128345)(0.252,0.128864)(0.2525,0.129383)(0.253,0.129904)(0.2535,0.130426)(0.254,0.130949)(0.2545,0.131473)(0.255,0.131998)(0.2555,0.132524)(0.256,0.133051)(0.2565,0.133579)(0.257,0.134109)(0.2575,0.134639)(0.258,0.135171)(0.2585,0.135703)(0.259,0.136237)(0.2595,0.136772)(0.26,0.137308)(0.2605,0.137845)(0.261,0.138383)(0.2615,0.138922)(0.262,0.139463)(0.2625,0.140004)(0.263,0.140547)(0.2635,0.141091)(0.264,0.141635)(0.2645,0.142181)(0.265,0.142728)(0.2655,0.143276)(0.266,0.143825)(0.2665,0.144376)(0.267,0.144927)(0.2675,0.14548)(0.268,0.146033)(0.2685,0.146588)(0.269,0.147144)(0.2695,0.147701)(0.27,0.148259)(0.2705,0.148818)(0.271,0.149378)(0.2715,0.14994)(0.272,0.150502)(0.2725,0.151066)(0.273,0.15163)(0.2735,0.152196)(0.274,0.152763)(0.2745,0.153331)(0.275,0.1539)(0.2755,0.15447)(0.276,0.155042)(0.2765,0.155614)(0.277,0.156188)(0.2775,0.156763)(0.278,0.157338)(0.2785,0.157915)(0.279,0.158493)(0.2795,0.159072)(0.28,0.159653)(0.2805,0.160234)(0.281,0.160817)(0.2815,0.1614)(0.282,0.161985)(0.2825,0.162571)(0.283,0.163158)(0.2835,0.163746)(0.284,0.164335)(0.2845,0.164926)(0.285,0.165517)(0.2855,0.16611)(0.286,0.166703)(0.2865,0.167298)(0.287,0.167894)(0.2875,0.168491)(0.288,0.169089)(0.2885,0.169689)(0.289,0.170289)(0.2895,0.170891)(0.29,0.171493)(0.2905,0.172097)(0.291,0.172702)(0.2915,0.173308)(0.292,0.173915)(0.2925,0.174524)(0.293,0.175133)(0.2935,0.175744)(0.294,0.176356)(0.2945,0.176968)(0.295,0.177582)(0.2955,0.178198)(0.296,0.178814)(0.2965,0.179431)(0.297,0.18005)(0.2975,0.180669)(0.298,0.18129)(0.2985,0.181912)(0.299,0.182535)(0.2995,0.183159)(0.3,0.183785)(0.3005,0.184411)(0.301,0.185039)(0.3015,0.185667)(0.302,0.186297)(0.3025,0.186928)(0.303,0.18756)(0.3035,0.188194)(0.304,0.188828)(0.3045,0.189464)(0.305,0.1901)(0.3055,0.190738)(0.306,0.191377)(0.3065,0.192017)(0.307,0.192659)(0.3075,0.193301)(0.308,0.193945)(0.3085,0.194589)(0.309,0.195235)(0.3095,0.195882)(0.31,0.19653)(0.3105,0.19718)(0.311,0.19783)(0.3115,0.198482)(0.312,0.199134)(0.3125,0.199788)(0.313,0.200443)(0.3135,0.201099)(0.314,0.201757)(0.3145,0.202415)(0.315,0.203075)(0.3155,0.203736)(0.316,0.204398)(0.3165,0.205061)(0.317,0.205725)(0.3175,0.20639)(0.318,0.207057)(0.3185,0.207725)(0.319,0.208394)(0.3195,0.209064)(0.32,0.209735)(0.3205,0.210407)(0.321,0.211081)(0.3215,0.211755)(0.322,0.212431)(0.3225,0.213108)(0.323,0.213786)(0.3235,0.214466)(0.324,0.215146)(0.3245,0.215828)(0.325,0.21651)(0.3255,0.217194)(0.326,0.21788)(0.3265,0.218566)(0.327,0.219253)(0.3275,0.219942)(0.328,0.220632)(0.3285,0.221323)(0.329,0.222015)(0.3295,0.222708)(0.33,0.223403)(0.3305,0.224098)(0.331,0.224795)(0.3315,0.225493)(0.332,0.226192)(0.3325,0.226892)(0.333,0.227594)(0.3335,0.228297)(0.334,0.229)(0.3345,0.229705)(0.335,0.230412)(0.3355,0.231119)(0.336,0.231827)(0.3365,0.232537)(0.337,0.233248)(0.3375,0.23396)(0.338,0.234673)(0.3385,0.235388)(0.339,0.236103)(0.3395,0.23682)(0.34,0.237538)(0.3405,0.238257)(0.341,0.238978)(0.3415,0.239699)(0.342,0.240422)(0.3425,0.241146)(0.343,0.241871)(0.3435,0.242597)(0.344,0.243325)(0.3445,0.244053)(0.345,0.244783)(0.3455,0.245514)(0.346,0.246246)(0.3465,0.24698)(0.347,0.247714)(0.3475,0.24845)(0.348,0.249187)(0.3485,0.249925)(0.349,0.250664)(0.3495,0.251405)(0.35,0.252147)(0.3505,0.25289)(0.351,0.253634)(0.3515,0.254379)(0.352,0.255126)(0.3525,0.255873)(0.353,0.256622)(0.3535,0.257372)(0.354,0.258124)(0.3545,0.258876)(0.355,0.25963)(0.3555,0.260385)(0.356,0.261141)(0.3565,0.261898)(0.357,0.262657)(0.3575,0.263417)(0.358,0.264178)(0.3585,0.26494)(0.359,0.265703)(0.3595,0.266468)(0.36,0.267233)(0.3605,0.268)(0.361,0.268769)(0.3615,0.269538)(0.362,0.270309)(0.3625,0.27108)(0.363,0.271853)(0.3635,0.272628)(0.364,0.273403)(0.3645,0.27418)(0.365,0.274958)(0.3655,0.275737)(0.366,0.276517)(0.3665,0.277299)(0.367,0.278081)(0.3675,0.278865)(0.368,0.27965)(0.3685,0.280437)(0.369,0.281225)(0.3695,0.282013)(0.37,0.282803)(0.3705,0.283595)(0.371,0.284387)(0.3715,0.285181)(0.372,0.285976)(0.3725,0.286772)(0.373,0.28757)(0.3735,0.288368)(0.374,0.289168)(0.3745,0.289969)(0.375,0.290772)(0.3755,0.291575)(0.376,0.29238)(0.3765,0.293186)(0.377,0.293993)(0.3775,0.294802)(0.378,0.295612)(0.3785,0.296423)(0.379,0.297235)(0.3795,0.298048)(0.38,0.298863)(0.3805,0.299679)(0.381,0.300496)(0.3815,0.301314)(0.382,0.302134)(0.3825,0.302955)(0.383,0.303777)(0.3835,0.3046)(0.384,0.305425)(0.3845,0.306251)(0.385,0.307078)(0.3855,0.307906)(0.386,0.308736)(0.3865,0.309567)(0.387,0.310399)(0.3875,0.311232)(0.388,0.312067)(0.3885,0.312903)(0.389,0.31374)(0.3895,0.314578)(0.39,0.315418)(0.3905,0.316258)(0.391,0.317101)(0.3915,0.317944)(0.392,0.318788)(0.3925,0.319634)(0.393,0.320481)(0.3935,0.32133)(0.394,0.32218)(0.3945,0.32303)(0.395,0.323883)(0.3955,0.324736)(0.396,0.325591)(0.3965,0.326447)(0.397,0.327304)(0.3975,0.328162)(0.398,0.329022)(0.3985,0.329883)(0.399,0.330746)(0.3995,0.331609)(0.4,0.332474)(0.4005,0.33334)(0.401,0.334207)(0.4015,0.335076)(0.402,0.335946)(0.4025,0.336817)(0.403,0.33769)(0.4035,0.338563)(0.404,0.339438)(0.4045,0.340315)(0.405,0.341192)(0.4055,0.342071)(0.406,0.342951)(0.4065,0.343833)(0.407,0.344715)(0.4075,0.345599)(0.408,0.346485)(0.4085,0.347371)(0.409,0.348259)(0.4095,0.349148)(0.41,0.350039)(0.4105,0.35093)(0.411,0.351823)(0.4115,0.352718)(0.412,0.353613)(0.4125,0.35451)(0.413,0.355408)(0.4135,0.356308)(0.414,0.357208)(0.4145,0.358111)(0.415,0.359014)(0.4155,0.359918)(0.416,0.360824)(0.4165,0.361732)(0.417,0.36264)(0.4175,0.36355)(0.418,0.364461)(0.4185,0.365374)(0.419,0.366287)(0.4195,0.367202)(0.42,0.368119)(0.4205,0.369036)(0.421,0.369955)(0.4215,0.370876)(0.422,0.371797)(0.4225,0.37272)(0.423,0.373644)(0.4235,0.37457)(0.424,0.375497)(0.4245,0.376425)(0.425,0.377354)(0.4255,0.378285)(0.426,0.379217)(0.4265,0.380151)(0.427,0.381086)(0.4275,0.382022)(0.428,0.382959)(0.4285,0.383898)(0.429,0.384838)(0.4295,0.385779)(0.43,0.386722)(0.4305,0.387666)(0.431,0.388611)(0.4315,0.389558)(0.432,0.390506)(0.4325,0.391455)(0.433,0.392406)(0.4335,0.393358)(0.434,0.394311)(0.4345,0.395266)(0.435,0.396222)(0.4355,0.397179)(0.436,0.398138)(0.4365,0.399098)(0.437,0.400059)(0.4375,0.401022)(0.438,0.401986)(0.4385,0.402951)(0.439,0.403918)(0.4395,0.404886)(0.44,0.405856)(0.4405,0.406826)(0.441,0.407798)(0.4415,0.408772)(0.442,0.409747)(0.4425,0.410723)(0.443,0.4117)(0.4435,0.412679)(0.444,0.413659)(0.4445,0.414641)(0.445,0.415624)(0.4455,0.416608)(0.446,0.417594)(0.4465,0.418581)(0.447,0.419569)(0.4475,0.420559)(0.448,0.42155)(0.4485,0.422543)(0.449,0.423536)(0.4495,0.424532)(0.45,0.425528)(0.4505,0.426526)(0.451,0.427525)(0.4515,0.428526)(0.452,0.429528)(0.4525,0.430531)(0.453,0.431536)(0.4535,0.432542)(0.454,0.43355)(0.4545,0.434559)(0.455,0.435569)(0.4555,0.436581)(0.456,0.437594)(0.4565,0.438608)(0.457,0.439624)(0.4575,0.440641)(0.458,0.44166)(0.4585,0.44268)(0.459,0.443701)(0.4595,0.444724)(0.46,0.445748)(0.4605,0.446774)(0.461,0.447801)(0.4615,0.448829)(0.462,0.449859)(0.4625,0.45089)(0.463,0.451923)(0.4635,0.452957)(0.464,0.453992)(0.4645,0.455029)(0.465,0.456067)(0.4655,0.457106)(0.466,0.458147)(0.4665,0.45919)(0.467,0.460233)(0.4675,0.461278)(0.468,0.462325)(0.4685,0.463373)(0.469,0.464422)(0.4695,0.465473)(0.47,0.466525)(0.4705,0.467579)(0.471,0.468634)(0.4715,0.469691)(0.472,0.470749)(0.4725,0.471808)(0.473,0.472869)(0.4735,0.473931)(0.474,0.474994)(0.4745,0.476059)(0.475,0.477126)(0.4755,0.478194)(0.476,0.479263)(0.4765,0.480334)(0.477,0.481406)(0.4775,0.48248)(0.478,0.483555)(0.4785,0.484631)(0.479,0.485709)(0.4795,0.486788)(0.48,0.487869)(0.4805,0.488951)(0.481,0.490035)(0.4815,0.49112)(0.482,0.492207)(0.4825,0.493295)(0.483,0.494384)(0.4835,0.495475)(0.484,0.496568)(0.4845,0.497661)(0.485,0.498757)(0.4855,0.499853)(0.486,0.500952)(0.4865,0.502051)(0.487,0.503152)(0.4875,0.504255)(0.488,0.505359)(0.4885,0.506464)(0.489,0.507571)(0.4895,0.50868)(0.49,0.50979)(0.4905,0.510901)(0.491,0.512014)(0.4915,0.513128)(0.492,0.514244)(0.4925,0.515361)(0.493,0.51648)(0.4935,0.5176)(0.494,0.518722)(0.4945,0.519845)(0.495,0.52097)(0.4955,0.522096)(0.496,0.523223)(0.4965,0.524352)(0.497,0.525483)(0.4975,0.526615)(0.498,0.527749)(0.4985,0.528884)(0.499,0.53002)(0.4995,0.531158)(0.5,0.532298)(0.5005,0.533439)(0.501,0.534581)(0.5015,0.535726)(0.502,0.536871)(0.5025,0.538018)(0.503,0.539167)(0.5035,0.540317)(0.504,0.541468)(0.5045,0.542621)(0.505,0.543776)(0.5055,0.544932)(0.506,0.54609)(0.5065,0.547249)(0.507,0.54841)(0.5075,0.549572)(0.508,0.550735)(0.5085,0.551901)(0.509,0.553067)(0.5095,0.554236)(0.51,0.555405)(0.5105,0.556577)(0.511,0.55775)(0.5115,0.558924)(0.512,0.5601)(0.5125,0.561277)(0.513,0.562456)(0.5135,0.563637)(0.514,0.564819)(0.5145,0.566002)(0.515,0.567188)(0.5155,0.568374)(0.516,0.569563)(0.5165,0.570752)(0.517,0.571944)(0.5175,0.573137)(0.518,0.574331)(0.5185,0.575527)(0.519,0.576725)(0.5195,0.577924)(0.52,0.579124)(0.5205,0.580327)(0.521,0.58153)(0.5215,0.582736)(0.522,0.583943)(0.5225,0.585151)(0.523,0.586361)(0.5235,0.587573)(0.524,0.588786)(0.5245,0.590001)(0.525,0.591217)(0.5255,0.592435)(0.526,0.593655)(0.5265,0.594876)(0.527,0.596098)(0.5275,0.597323)(0.528,0.598548)(0.5285,0.599776)(0.529,0.601005)(0.5295,0.602235)(0.53,0.603468)(0.5305,0.604701)(0.531,0.605937)(0.5315,0.607174)(0.532,0.608412)(0.5325,0.609653)(0.533,0.610894)(0.5335,0.612138)(0.534,0.613383)(0.5345,0.61463)(0.535,0.615878)(0.5355,0.617128)(0.536,0.618379)(0.5365,0.619632)(0.537,0.620887)(0.5375,0.622143)(0.538,0.623401)(0.5385,0.624661)(0.539,0.625922)(0.5395,0.627185)(0.54,0.628449)(0.5405,0.629715)(0.541,0.630983)(0.5415,0.632252)(0.542,0.633523)(0.5425,0.634796)(0.543,0.63607)(0.5435,0.637346)(0.544,0.638624)(0.5445,0.639903)(0.545,0.641184)(0.5455,0.642466)(0.546,0.643751)(0.5465,0.645036)(0.547,0.646324)(0.5475,0.647613)(0.548,0.648904)(0.5485,0.650196)(0.549,0.65149)(0.5495,0.652786)(0.55,0.654084)(0.5505,0.655383)(0.551,0.656684)(0.5515,0.657986)(0.552,0.65929)(0.5525,0.660596)(0.553,0.661904)(0.5535,0.663213)(0.554,0.664524)(0.5545,0.665836)(0.555,0.667151)(0.5555,0.668467)(0.556,0.669784)(0.5565,0.671104)(0.557,0.672425)(0.5575,0.673747)(0.558,0.675072)(0.5585,0.676398)(0.559,0.677726)(0.5595,0.679055)(0.56,0.680387)(0.5605,0.68172)(0.561,0.683054)(0.5615,0.684391)(0.562,0.685729)(0.5625,0.687069)(0.563,0.68841)(0.5635,0.689754)(0.564,0.691099)(0.5645,0.692445)(0.565,0.693794)(0.5655,0.695144)(0.566,0.696496)(0.5665,0.69785)(0.567,0.699205)(0.5675,0.700563)(0.568,0.701922)(0.5685,0.703282)(0.569,0.704645)(0.5695,0.706009)(0.57,0.707375)(0.5705,0.708743)(0.571,0.710112)(0.5715,0.711483)(0.572,0.712856)(0.5725,0.714231)(0.573,0.715608)(0.5735,0.716986)(0.574,0.718366)(0.5745,0.719748)(0.575,0.721131)(0.5755,0.722517)(0.576,0.723904)(0.5765,0.725293)(0.577,0.726684)(0.5775,0.728076)(0.578,0.729471)(0.5785,0.730867)(0.579,0.732265)(0.5795,0.733664)(0.58,0.735066)(0.5805,0.736469)(0.581,0.737875)(0.5815,0.739282)(0.582,0.74069)(0.5825,0.742101)(0.583,0.743513)(0.5835,0.744928)(0.584,0.746344)(0.5845,0.747761)(0.585,0.749181)(0.5855,0.750603)(0.586,0.752026)(0.5865,0.753451)(0.587,0.754878)(0.5875,0.756307)(0.588,0.757738)(0.5885,0.759171)(0.589,0.760605)(0.5895,0.762041)(0.59,0.763479)(0.5905,0.764919)(0.591,0.766361)(0.5915,0.767805)(0.592,0.76925)(0.5925,0.770698)(0.593,0.772147)(0.5935,0.773598)(0.594,0.775051)(0.5945,0.776506)(0.595,0.777963)(0.5955,0.779422)(0.596,0.780882)(0.5965,0.782345)(0.597,0.783809)(0.5975,0.785276)(0.598,0.786744)(0.5985,0.788214)(0.599,0.789686)(0.5995,0.79116)(0.6,0.792635)(0.6005,0.794113)(0.601,0.795593)(0.6015,0.797074)(0.602,0.798557)(0.6025,0.800043)(0.603,0.80153)(0.6035,0.803019)(0.604,0.80451)(0.6045,0.806004)(0.605,0.807499)(0.6055,0.808995)(0.606,0.810494)(0.6065,0.811995)(0.607,0.813498)(0.6075,0.815003)(0.608,0.816509)(0.6085,0.818018)(0.609,0.819529)(0.6095,0.821041)(0.61,0.822556)(0.6105,0.824072)(0.611,0.825591)(0.6115,0.827111)(0.612,0.828634)(0.6125,0.830158)(0.613,0.831685)(0.6135,0.833213)(0.614,0.834743)(0.6145,0.836276)(0.615,0.83781)(0.6155,0.839347)(0.616,0.840885)(0.6165,0.842425)(0.617,0.843968)(0.6175,0.845512)(0.618,0.847059)(0.6185,0.848607)(0.619,0.850158)(0.6195,0.85171)(0.62,0.853265)(0.6205,0.854821)(0.621,0.85638)(0.6215,0.857941)(0.622,0.859503)(0.6225,0.861068)(0.623,0.862635)(0.6235,0.864204)(0.624,0.865775)(0.6245,0.867348)(0.625,0.868923)(0.6255,0.8705)(0.626,0.872079)(0.6265,0.87366)(0.627,0.875243)(0.6275,0.876829)(0.628,0.878416)(0.6285,0.880006)(0.629,0.881598)(0.6295,0.883191)(0.63,0.884787)(0.6305,0.886385)(0.631,0.887985)(0.6315,0.889588)(0.632,0.891192)(0.6325,0.892798)(0.633,0.894407)(0.6335,0.896018)(0.634,0.89763)(0.6345,0.899245)(0.635,0.900862)(0.6355,0.902482)(0.636,0.904103)(0.6365,0.905727)(0.637,0.907352)(0.6375,0.90898)(0.638,0.91061)(0.6385,0.912242)(0.639,0.913876)(0.6395,0.915513)(0.64,0.917152)(0.6405,0.918792)(0.641,0.920435)(0.6415,0.922081)(0.642,0.923728)(0.6425,0.925378)(0.643,0.927029)(0.6435,0.928683)(0.644,0.930339)(0.6445,0.931998)(0.645,0.933658)(0.6455,0.935321)(0.646,0.936986)(0.6465,0.938654)(0.647,0.940323)(0.6475,0.941995)(0.648,0.943669)(0.6485,0.945345)(0.649,0.947024)(0.6495,0.948704)(0.65,0.950387)(0.6505,0.952072)(0.651,0.95376)(0.6515,0.95545)(0.652,0.957142)(0.6525,0.958836)(0.653,0.960533)(0.6535,0.962232)(0.654,0.963933)(0.6545,0.965636)(0.655,0.967342)(0.6555,0.96905)(0.656,0.97076)(0.6565,0.972473)(0.657,0.974188)(0.6575,0.975905)(0.658,0.977625)(0.6585,0.979347)(0.659,0.981071)(0.6595,0.982798)(0.66,0.984527)(0.6605,0.986258)(0.661,0.987992)(0.6615,0.989728)(0.662,0.991467)(0.6625,0.993207)(0.663,0.994951)(0.6635,0.996696)(0.664,0.998444)(0.6645,1.00019)(0.665,1.00195)(0.6655,1.0037)(0.666,1.00546)(0.6665,1.00722)(0.667,1.00898)(0.6675,1.01075)(0.668,1.01251)(0.6685,1.01428)(0.669,1.01606)(0.6695,1.01783)(0.67,1.01961)(0.6705,1.02139)(0.671,1.02317)(0.6715,1.02495)(0.672,1.02674)(0.6725,1.02853)(0.673,1.03032)(0.6735,1.03211)(0.674,1.03391)(0.6745,1.03571)(0.675,1.03751)(0.6755,1.03931)(0.676,1.04112)(0.6765,1.04293)(0.677,1.04474)(0.6775,1.04656)(0.678,1.04837)(0.6785,1.05019)(0.679,1.05201)(0.6795,1.05384)(0.68,1.05566)(0.6805,1.05749)(0.681,1.05933)(0.6815,1.06116)(0.682,1.063)(0.6825,1.06484)(0.683,1.06668)(0.6835,1.06852)(0.684,1.07037)(0.6845,1.07222)(0.685,1.07407)(0.6855,1.07593)(0.686,1.07779)(0.6865,1.07965)(0.687,1.08151)(0.6875,1.08338)(0.688,1.08524)(0.6885,1.08712)(0.689,1.08899)(0.6895,1.09087)(0.69,1.09274)(0.6905,1.09463)(0.691,1.09651)(0.6915,1.0984)(0.692,1.10029)(0.6925,1.10218)(0.693,1.10407)(0.6935,1.10597)(0.694,1.10787)(0.6945,1.10978)(0.695,1.11168)(0.6955,1.11359)(0.696,1.1155)(0.6965,1.11742)(0.697,1.11933)(0.6975,1.12125)(0.698,1.12317)(0.6985,1.1251)(0.699,1.12703)(0.6995,1.12896)(0.7,1.13089)(0.7005,1.13283)(0.701,1.13477)(0.7015,1.13671)(0.702,1.13865)(0.7025,1.1406)(0.703,1.14255)(0.7035,1.14451)(0.704,1.14646)(0.7045,1.14842)(0.705,1.15038)(0.7055,1.15235)(0.706,1.15431)(0.7065,1.15629)(0.707,1.15826)(0.7075,1.16024)(0.708,1.16221)(0.7085,1.1642)(0.709,1.16618)(0.7095,1.16817)(0.71,1.17016)(0.7105,1.17215)(0.711,1.17415)(0.7115,1.17615)(0.712,1.17815)(0.7125,1.18016)(0.713,1.18217)(0.7135,1.18418)(0.714,1.18619)(0.7145,1.18821)(0.715,1.19023)(0.7155,1.19226)(0.716,1.19428)(0.7165,1.19631)(0.717,1.19835)(0.7175,1.20038)(0.718,1.20242)(0.7185,1.20446)(0.719,1.20651)(0.7195,1.20856)(0.72,1.21061)(0.7205,1.21266)(0.721,1.21472)(0.7215,1.21678)(0.722,1.21884)(0.7225,1.22091)(0.723,1.22298)(0.7235,1.22505)(0.724,1.22713)(0.7245,1.22921)(0.725,1.23129)(0.7255,1.23338)(0.726,1.23547)(0.7265,1.23756)(0.727,1.23965)(0.7275,1.24175)(0.728,1.24386)(0.7285,1.24596)(0.729,1.24807)(0.7295,1.25018)(0.73,1.2523)(0.7305,1.25441)(0.731,1.25654)(0.7315,1.25866)(0.732,1.26079)(0.7325,1.26292)(0.733,1.26506)(0.7335,1.26719)(0.734,1.26934)(0.7345,1.27148)(0.735,1.27363)(0.7355,1.27578)(0.736,1.27794)(0.7365,1.28009)(0.737,1.28226)(0.7375,1.28442)(0.738,1.28659)(0.7385,1.28876)(0.739,1.29094)(0.7395,1.29312)(0.74,1.2953)(0.7405,1.29749)(0.741,1.29968)(0.7415,1.30187)(0.742,1.30407)(0.7425,1.30627)(0.743,1.30847)(0.7435,1.31068)(0.744,1.31289)(0.7445,1.3151)(0.745,1.31732)(0.7455,1.31954)(0.746,1.32177)(0.7465,1.324)(0.747,1.32623)(0.7475,1.32847)(0.748,1.33071)(0.7485,1.33295)(0.749,1.3352)(0.7495,1.33745)(0.75,1.3397)(0.7505,1.34196)(0.751,1.34422)(0.7515,1.34649)(0.752,1.34876)(0.7525,1.35103)(0.753,1.35331)(0.7535,1.35559)(0.754,1.35787)(0.7545,1.36016)(0.755,1.36246)(0.7555,1.36475)(0.756,1.36705)(0.7565,1.36936)(0.757,1.37166)(0.7575,1.37398)(0.758,1.37629)(0.7585,1.37861)(0.759,1.38093)(0.7595,1.38326)(0.76,1.38559)(0.7605,1.38793)(0.761,1.39027)(0.7615,1.39261)(0.762,1.39496)(0.7625,1.39731)(0.763,1.39967)(0.7635,1.40203)(0.764,1.40439)(0.7645,1.40676)(0.765,1.40913)(0.7655,1.41151)(0.766,1.41389)(0.7665,1.41627)(0.767,1.41866)(0.7675,1.42105)(0.768,1.42345)(0.7685,1.42585)(0.769,1.42826)(0.7695,1.43067)(0.77,1.43308)(0.7705,1.4355)(0.771,1.43792)(0.7715,1.44035)(0.772,1.44278)(0.7725,1.44521)(0.773,1.44765)(0.7735,1.4501)(0.774,1.45255)(0.7745,1.455)(0.775,1.45746)(0.7755,1.45992)(0.776,1.46239)(0.7765,1.46486)(0.777,1.46733)(0.7775,1.46981)(0.778,1.4723)(0.7785,1.47479)(0.779,1.47728)(0.7795,1.47978)(0.78,1.48228)(0.7805,1.48479)(0.781,1.4873)(0.7815,1.48982)(0.782,1.49234)(0.7825,1.49487)(0.783,1.4974)(0.7835,1.49994)(0.784,1.50248)(0.7845,1.50502)(0.785,1.50757)(0.7855,1.51013)(0.786,1.51269)(0.7865,1.51525)(0.787,1.51782)(0.7875,1.5204)(0.788,1.52298)(0.7885,1.52556)(0.789,1.52815)(0.7895,1.53075)(0.79,1.53335)(0.7905,1.53595)(0.791,1.53856)(0.7915,1.54118)(0.792,1.5438)(0.7925,1.54642)(0.793,1.54905)(0.7935,1.55169)(0.794,1.55433)(0.7945,1.55697)(0.795,1.55962)(0.7955,1.56228)(0.796,1.56494)(0.7965,1.56761)(0.797,1.57028)(0.7975,1.57296)(0.798,1.57564)(0.7985,1.57833)(0.799,1.58103)(0.7995,1.58373)(0.8,1.58643)(0.8005,1.58914)(0.801,1.59186)(0.8015,1.59458)(0.802,1.59731)(0.8025,1.60004)(0.803,1.60278)(0.8035,1.60552)(0.804,1.60827)(0.8045,1.61103)(0.805,1.61379)(0.8055,1.61656)(0.806,1.61933)(0.8065,1.62211)(0.807,1.6249)(0.8075,1.62769)(0.808,1.63049)(0.8085,1.63329)(0.809,1.6361)(0.8095,1.63891)(0.81,1.64174)(0.8105,1.64456)(0.811,1.6474)(0.8115,1.65024)(0.812,1.65308)(0.8125,1.65593)(0.813,1.65879)(0.8135,1.66166)(0.814,1.66453)(0.8145,1.66741)(0.815,1.67029)(0.8155,1.67318)(0.816,1.67608)(0.8165,1.67898)(0.817,1.68189)(0.8175,1.68481)(0.818,1.68773)(0.8185,1.69066)(0.819,1.6936)(0.8195,1.69654)(0.82,1.69949)(0.8205,1.70245)(0.821,1.70541)(0.8215,1.70838)(0.822,1.71136)(0.8225,1.71434)(0.823,1.71733)(0.8235,1.72033)(0.824,1.72334)(0.8245,1.72635)(0.825,1.72937)(0.8255,1.7324)(0.826,1.73543)(0.8265,1.73847)(0.827,1.74152)(0.8275,1.74458)(0.828,1.74764)(0.8285,1.75071)(0.829,1.75379)(0.8295,1.75688)(0.83,1.75997)(0.8305,1.76307)(0.831,1.76618)(0.8315,1.7693)(0.832,1.77243)(0.8325,1.77556)(0.833,1.7787)(0.8335,1.78185)(0.834,1.785)(0.8345,1.78817)(0.835,1.79134)(0.8355,1.79452)(0.836,1.79771)(0.8365,1.80091)(0.837,1.80412)(0.8375,1.80733)(0.838,1.81055)(0.8385,1.81378)(0.839,1.81702)(0.8395,1.82027)(0.84,1.82353)(0.8405,1.82679)(0.841,1.83007)(0.8415,1.83335)(0.842,1.83664)(0.8425,1.83994)(0.843,1.84325)(0.8435,1.84657)(0.844,1.8499)(0.8445,1.85324)(0.845,1.85658)(0.8455,1.85994)(0.846,1.8633)(0.8465,1.86668)(0.847,1.87006)(0.8475,1.87346)(0.848,1.87686)(0.8485,1.88027)(0.849,1.88369)(0.8495,1.88713)(0.85,1.89057)(0.8505,1.89402)(0.851,1.89748)(0.8515,1.90096)(0.852,1.90444)(0.8525,1.90793)(0.853,1.91144)(0.8535,1.91495)(0.854,1.91848)(0.8545,1.92201)(0.855,1.92556)(0.8555,1.92911)(0.856,1.93268)(0.8565,1.93626)(0.857,1.93985)(0.8575,1.94345)(0.858,1.94706)(0.8585,1.95069)(0.859,1.95432)(0.8595,1.95797)(0.86,1.96162)(0.8605,1.96529)(0.861,1.96897)(0.8615,1.97267)(0.862,1.97637)(0.8625,1.98009)(0.863,1.98382)(0.8635,1.98756)(0.864,1.99131)(0.8645,1.99507)(0.865,1.99885)(0.8655,2.00264)(0.866,2.00645)(0.8665,2.01026)(0.867,2.01409)(0.8675,2.01793)(0.868,2.02179)(0.8685,2.02566)(0.869,2.02954)(0.8695,2.03344)(0.87,2.03735)(0.8705,2.04127)(0.871,2.04521)(0.8715,2.04916)(0.872,2.05312)(0.8725,2.0571)(0.873,2.06109)(0.8735,2.0651)(0.874,2.06912)(0.8745,2.07316)(0.875,2.07721)(0.8755,2.08128)(0.876,2.08536)(0.8765,2.08946)(0.877,2.09357)(0.8775,2.0977)(0.878,2.10184)(0.8785,2.106)(0.879,2.11018)(0.8795,2.11437)(0.88,2.11858)(0.8805,2.1228)(0.881,2.12704)(0.8815,2.1313)(0.882,2.13557)(0.8825,2.13986)(0.883,2.14417)(0.8835,2.1485)(0.884,2.15284)(0.8845,2.1572)(0.885,2.16158)(0.8855,2.16598)(0.886,2.17039)(0.8865,2.17483)(0.887,2.17928)(0.8875,2.18375)(0.888,2.18824)(0.8885,2.19275)(0.889,2.19727)(0.8895,2.20182)(0.89,2.20639)(0.8905,2.21098)(0.891,2.21558)(0.8915,2.22021)(0.892,2.22486)(0.8925,2.22953)(0.893,2.23421)(0.8935,2.23893)(0.894,2.24366)(0.8945,2.24841)(0.895,2.25319)(0.8955,2.25798)(0.896,2.2628)(0.8965,2.26764)(0.897,2.27251)(0.8975,2.27739)(0.898,2.2823)(0.8985,2.28724)(0.899,2.2922)(0.8995,2.29718)(0.9,2.30218)(0.9005,2.30721)(0.901,2.31227)(0.9015,2.31735)(0.902,2.32245)(0.9025,2.32758)(0.903,2.33274)(0.9035,2.33792)(0.904,2.34313)(0.9045,2.34836)(0.905,2.35363)(0.9055,2.35891)(0.906,2.36423)(0.9065,2.36958)(0.907,2.37495)(0.9075,2.38035)(0.908,2.38578)(0.9085,2.39124)(0.909,2.39673)(0.9095,2.40225)(0.91,2.4078)(0.9105,2.41337)(0.911,2.41898)(0.9115,2.42463)(0.912,2.4303)(0.9125,2.436)(0.913,2.44174)(0.9135,2.44751)(0.914,2.45331)(0.9145,2.45915)(0.915,2.46502)(0.9155,2.47093)(0.916,2.47687)(0.9165,2.48284)(0.917,2.48885)(0.9175,2.4949)(0.918,2.50098)(0.9185,2.5071)(0.919,2.51326)(0.9195,2.51945)(0.92,2.52569)(0.9205,2.53196)(0.921,2.53827)(0.9215,2.54462)(0.922,2.55101)(0.9225,2.55745)(0.923,2.56392)(0.9235,2.57044)(0.924,2.577)(0.9245,2.5836)(0.925,2.59025)(0.9255,2.59694)(0.926,2.60367)(0.9265,2.61045)(0.927,2.61728)(0.9275,2.62416)(0.928,2.63108)(0.9285,2.63805)(0.929,2.64507)(0.9295,2.65213)(0.93,2.65925)(0.9305,2.66642)(0.931,2.67364)(0.9315,2.68092)(0.932,2.68824)(0.9325,2.69562)(0.933,2.70306)(0.9335,2.71055)(0.934,2.7181)(0.9345,2.7257)(0.935,2.73337)(0.9355,2.74109)(0.936,2.74887)(0.9365,2.75671)(0.937,2.76462)(0.9375,2.77259)(0.938,2.78062)(0.9385,2.78872)(0.939,2.79688)(0.9395,2.80511)(0.94,2.81341)(0.9405,2.82178)(0.941,2.83022)(0.9415,2.83873)(0.942,2.84731)(0.9425,2.85597)(0.943,2.8647)(0.9435,2.87351)(0.944,2.8824)(0.9445,2.89137)(0.945,2.90042)(0.9455,2.90955)(0.946,2.91877)(0.9465,2.92807)(0.947,2.93746)(0.9475,2.94694)(0.948,2.95651)(0.9485,2.96617)(0.949,2.97593)(0.9495,2.98578)(0.95,2.99573)(0.9505,3.00578)(0.951,3.01593)(0.9515,3.02619)(0.952,3.03655)(0.9525,3.04703)(0.953,3.05761)(0.9535,3.0683)(0.954,3.07911)(0.9545,3.09004)(0.955,3.10109)(0.9555,3.11227)(0.956,3.12357)(0.9565,3.13499)(0.957,3.14656)(0.9575,3.15825)(0.958,3.17009)(0.9585,3.18206)(0.959,3.19418)(0.9595,3.20645)(0.96,3.21888)(0.9605,3.23145)(0.961,3.24419)(0.9615,3.2571)(0.962,3.27017)(0.9625,3.28341)(0.963,3.29684)(0.9635,3.31044)(0.964,3.32424)(0.9645,3.33822)(0.965,3.35241)(0.9655,3.3668)(0.966,3.38139)(0.9665,3.39621)(0.967,3.41125)(0.9675,3.42652)(0.968,3.44202)(0.9685,3.45777)(0.969,3.47377)(0.9695,3.49003)(0.97,3.50656)(0.9705,3.52337)(0.971,3.54046)(0.9715,3.55785)(0.972,3.57555)(0.9725,3.59357)(0.973,3.61192)(0.9735,3.63061)(0.974,3.64966)(0.9745,3.66908)(0.975,3.68888)(0.9755,3.70908)(0.976,3.7297)(0.9765,3.75075)(0.977,3.77226)(0.9775,3.79424)(0.978,3.81671)(0.9785,3.8397)(0.979,3.86323)(0.9795,3.88733)(0.98,3.91202)(0.9805,3.93734)(0.981,3.96332)(0.9815,3.98998)(0.982,4.01738)(0.9825,4.04555)(0.983,4.07454)(0.9835,4.10439)(0.984,4.13517)(0.9845,4.16692)(0.985,4.19971)(0.9855,4.23361)(0.986,4.2687)(0.9865,4.30507)(0.987,4.34281)(0.9875,4.38203)(0.988,4.42285)(0.9885,4.46541)(0.989,4.50986)(0.9895,4.55638)(0.99,4.60517)(0.9905,4.65646)(0.991,4.71053)(0.9915,4.76769)(0.992,4.82831)(0.9925,4.89285)(0.993,4.96185)(0.9935,5.03595)(0.994,5.116)(0.9945,5.20301)(0.995,5.29832)(0.9955,5.40368)(0.996,5.52146)(0.9965,5.65499)(0.997,5.80914)(0.9975,5.99146)(0.998,6.21461)(0.9985,6.50229)(0.999,6.90776)};
       \addlegendentry{$h^*$ (implicit)};
\addplot[samples = 200, smooth,thick,black]  {x*ln((1+x)/(1-x))};
  \addlegendentry{$h^*_\text{sym}$};
\end{axis}
\end{tikzpicture}
\caption{
The function $h(x)$ as given by Pinsker, Gilardoni, the optimal $h^*$ given (in implicit  form) by Fedotov et al. \cite{fedotov2003refinements}, and the tightest bound $h^*_\text{sym}$ on the symmetric Kullback-Leibler divergence.}
\label{fig:bounds}
\end{figure}
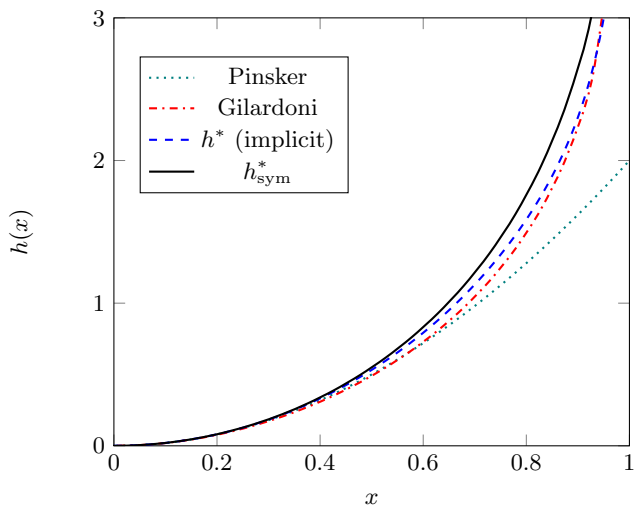

The situation is simpler for a symmetric version of Kullback-Leibler divergences, as we can write:

\begin{align}\label{eq:Dpqhsym}
	\frac{D(p\|q)+D(q\|p)}{2} \geq h_\text{sym}(d_{TV}(p,q)).	
\end{align}
for some convex increasing function $h_\text{sym}$. Of course we can take any function $h$ as above, for instance Pinsker's $h_\text{sym}(x)=2x^2$. Interestingly however, none of these are tight, and the tighest bound has an explicit formulation:

\begin{align} \label{eq:hstarsym}
	h^*_\text{sym}(x)= x \ln \frac{1+x}{1-x}=2x\, \text{atanh}\, x,	
\end{align}
as proved e.g. in \cite{gilardoni2008improvement} and \cite{falasco2021beyond}.
It is verified with equality for the two-element probability distributions $p = (\epsilon, 1-\epsilon)$ and $q=(1-\epsilon, \epsilon)$. In fact, it is more generally tight for any distributions $p$ and $q$ so that $|\ln p(\omega)/q(\omega)| = c$, for some constant $c >0$ and all $\omega \in \Omega$. See Fig. \eqref{fig:bounds} for a graphical comparison of these bounds.

\subsection{Entropy production and its decompositions} In this section we recall some elementary concepts of the stochastic thermodynamics of Markov processes, written here in the language of Kullback-Leibler divergences.
The goal of this section is to establish the equalities \eqref{eq:sigmaNAHH}, \eqref{eq:overlinevstilde} and \eqref{eq:Dpptildeissym} relating various concepts of entropy production, along with the necessary notations.

 The entropy production of a time-varying Markov process over a discrete state space over a given time interval $[\tau_\text{ini},\tau_\text{fin}]$ is defined as  
\begin{align}
\Delta \sigma &=  D(p\|\overline{p})=   \sum_{\omega} p(\omega)  \ln \frac{p(\omega)}{\overline{p}(\omega)},
\end{align}
where $\omega$ runs over all trajectories of the Markov process on the time interval. Here the time-reversed probability distribution $\overline{p}(\omega)$ denotes the probability of the time-reversed trajectory $\overline{\omega}$  in the time-reversed process, whose initial state distribution is defined to be the final state distribution predicted by $p$, and the transition rates are defined as the transition rates of $p$, read backward in time (in case these rates are time-varying).  We assume here an ``overdamped'' framework, where the time-reversed trajectory is simply the list of states and transitions read in reverse order. %See Discussion for adaptation to the ``underdamped'' case, namely, when some variables flip sign under time reversal.

Whenever  $\Omega$ is the set of trajectories of a Markov process taking place in a state space $\mathcal{X}$, we define two observables $X_\text{ini}$, the initial state of the trajectory, and $X_\text{fin}$ the final state of the trajetory.   By construction of the time-reversed  process $\overline{p}$ we have in particular:
\begin{align}
\overline{p}_{X_\text{fin}}(x)&=\sum_{\omega:  X_\text{fin}(\omega)=x} \overline{p}(\omega)\\
		&=\sum_{\omega:  X_\text{ini}(\overline{\omega})=x} \overline{p}(\omega).
\end{align}
  Since $\overline{p}(\omega)$ computes the probability of $\overline{\omega}$ starting with initial distribution $p_{X_\text{fin}}$, we see by definition
\begin{align} \label{eq:finfin}
 p_{X_\text{fin}}=\overline{p}_{X_\text{fin}}.	
\end{align}	

When the transition rates are \emph{constant} over time, or more generally \emph{symmetric under time-reversal}, it is sometimes convenient to consider, as an intermediate quantity, the probability distribution $\tilde{p}(\omega)=p(\overline{\omega})$: the trajectory is reversed but not the   process (i.e.,  we keep the same initial distribution and the same sequence of rates as in $p$).    In particular:
\begin{align}
	\tilde{p}_{X_\text{fin}}(x)&=\sum_{\omega:  X_\text{fin}(\omega)=x} \tilde{p}(\omega)\\
	&=\sum_{\omega:  X_\text{fin}(\omega)=x} p(\overline{\omega})\\
	&=\sum_{\omega:  X_\text{ini}(\overline{\omega})=x} p(\overline{\omega})\\
	&=p_{X_\text{ini}}.
\end{align} In summary we can write, in contrast to \eqref{eq:finfin}:
\begin{align}
	p_{X_\text{ini}}=\tilde{p}_{X_\text{fin}}.	
\end{align}

Since the rates are constant or time-symmetric,    $\tilde{p}(\omega)$ and $\overline{p}(\omega)$ compute the probability of $\overline{\omega}$ with the same transition rates but with a different probability for the initial state of $\overline{\omega}$ (which is the final state of $\omega$), which are $p_{X_\text{ini}}$ and $p_{X_\text{fin}}$ respectively.

Overall we can write, for constant-rate or time-symmetric-rate Markov chains :

\begin{align}
	D(p\|\overline{p})&=   \sum_{\omega} p(\omega)  \ln \frac{p(\omega)}{\tilde{p}(\omega)}\frac{\tilde{p}(\omega)}{\overline{p}(\omega)} \\&= D(p\|\tilde{p})+\sum_{\omega} p(\omega) \ln \frac{p_{X_\text{ini}(X_\text{fin}(\omega))}}{p_{X_\text{fin}(X_\text{fin}(\omega))}}\\
	&=D(p\|\tilde{p})- D(p_{X_\text{fin}}\|p_{X_\text{ini}}) \\
	&\leq D(p\|\tilde{p}).
\end{align}
Equality holds for stationary Markov chains,   { or for infinitesimal time intervals, as over a very short interval $[\tau_\text{ini},\tau_\text{fin}]$, $D(p_{X_\text{fin}}\|p_{X_\text{ini}})$  is of order $(\tau_\text{fin} -\tau_\text{ini})^2$ (the square Fisher metric), thus is negligible before a nonzero $\Delta \sigma$ (which is typically in the order of $\tau_\text{fin} -\tau_\text{ini}$).}

Considering two distributions $p$ and $q$ on the trajectory space $\Omega$ of some Markov process (possibly with time-varying transition rates), which differ only in their initial distribution (obeying the same transition rates) one can write the Chain Rule for the initial state, finding
\begin{align}
	D(p\|q)=D(p_{X_\text{ini}}\|q_{X_\text{ini}})+D(p\|q|X_\text{ini})=D(p_{X_\text{ini}}\|q_{X_\text{ini}}),
\end{align}
and for the final state,
\begin{align}
D(p\|q)=D(p_{X_\text{fin}}\|q_{X_\text{fin}})+D(p\|q|X_\text{fin}).
\end{align}
Comparing the two equations we find
\begin{align}
	D(p_{X_\text{ini}}\|q_{X_\text{ini}})-D(p_{X_\text{fin}}\|q_{X_\text{fin}}) = D(p\|q|X_\text{fin}) \geq 0.
\end{align}
This is one standard way \cite{cover1999elements} to show that the Kullback-Leibler divergence between two state distributions driven by the same rates is non-increasing. We can write further:

\begin{align}
	D(p\|\overline{p})&=D(p\|\overline{p} | X_\text{fin}) + D(p_{X_\text{fin}}  \| \overline{p}_{X_\text{fin}} )\\ 
	&=D(p\|\overline{q} |  X_\text{fin}) \\
	&=D(p\|q |  X_\text{fin}) +(D(p\|\overline{q} |  X_\text{fin}) - D(p\|q |  X_\text{fin})  ) \nonumber \\
	&=D(p\|q |  X_\text{fin}) +(D(p\|\overline{q} ) - D(p\|q )  ) \label{eq:NAplusHH}
\end{align}

We have used \eqref{eq:finfin}, the fact that $p_{X_\text{fin}} = \overline{p}_{X_\text{fin}}$, the fact that $\overline{q}(\omega)$ and $\overline{p}(\omega)$ are identical conditional to the final state of $\omega$ (as they obey the same time-varying rates), and the fact that $\overline{q}_{X_\text{fin}}=q_{X_\text{fin}}$.

Although this decomposition is true for any $q$ obeying the same time-varying rates as $p$, it takes a particularly significant meaning when rates are constant in time, or time-symmetric, and $q$ is a stationary process obeying those rates ($q_{X_\text{fin}}=q_{X_\text{ini}}$, thus $\overline{q}=\tilde{q}$).
In this case  $D(p\|q|X_\text{fin})$ is called the \emph{non-adiabatic entropy production} $\Delta \sigma_{NA}$  \cite{esposito2010three} or excess entropy production \cite{oono1998steady} or Hatano-Sasa entropy production \cite{hat01}, and is zero if and only if $p$ is also stationary.

 Furthermore $D(p\|\overline{q} ) - D(p\|q )=D(p\|\overline{q} | X_\text{ini} ) \geq 0$ is called the \emph{housekeeping entropy production} $\Delta \sigma_{HK}$, or adiabatic entropy production \cite{esposito2010three}, which is zero if and only if detailed balance is satisfied, i.e., if the $q$ is an equilibrium, i.e. time-reversible ($q=\overline{q}$), i.e. driven by conservative forces.   

In summary we have proved the following decomposition \cite{esposito2010three} for entropy production $\Delta \sigma= D(p\|\overline{p})$ on a constant-rate or time-symmetric-rate Markov process with  $q$ as the corresponding stationary process:
\begin{align} \label{eq:sigmaNAHH}
	\Delta \sigma &= \Delta \sigma_{NA} + \Delta \sigma_{HK}\\
	 			&=  D(p\|q |  X_\text{fin})+ D(p\|\overline{q} | X_\text{ini} )  	 	.		
\end{align}
In addition, we have proved, in the same context,
\begin{align}\label{eq:overlinevstilde}
	\Delta \sigma= D(p\|\tilde{p}) - D(p_{X_\text{fin}}\|p_{X_\text{ini}}).
\end{align}

Finally, another well-known observation is  that for any set of trajectories $\Omega$, we find  
\begin{align}
	D(p\|\tilde{p})  &= \frac{1}{2}  \sum_{\omega} (p(\omega) -p(\overline{\omega})) \ln \frac{p(\omega)}{p(\overline{\omega})}\\
	&=\frac{D(p\|\tilde{p}) + D(\tilde{p}\|p)}{2}. \label{eq:Dpptildeissym}
\end{align}
To evaluate the entropy productions in case of a general continuous-time discrete-state Markov chain with  time-varying rates, one can decompose the total time interval into successive infinitesimal time intervals $\tau_\text{fin}-\tau_\text{ini}=dt$, where the rates can be considered constant, and sum the entropy production computed over each interval. Note that in this process, the terms $D(p_{X_\text{fin}}\|p_{X_\text{ini}})$, of the negligible order  $\mathcal{O}(dt^2)$, vanish. Recall as well that in an interval $dt$, the trajectory $\omega$ of a Markov jump process either is constant (no transition, with probability $1-\mathcal{O}(dt)$), or consists in a single transition between initial and final state (with probability $\mathcal{O}(dt)$). We can safely neglect the possibility of two or more transitions in an interval $dt$, as it occurs with negligible probability $\mathcal{O}(dt^2)$. 

\section{Thermo-Kinetic Relations for entropy production}

\label{sec:2}

 \subsection{Constant or time-symmetric rates}
  Let $A$ be a time-symmetric nonnegative function on trajectories in $\Omega$. By time-symmetric, we mean symmetric under time reversal, i.e. $A(\omega)=A(\overline{\omega})$ for all trajectories $\omega$.
Thus, for a constant-rate or time-symmetric-rate Markov process over arbitrary time intervals, we find, following \eqref{eq:overlinevstilde},\eqref{eq:Dpptildeissym},\eqref{eq:DpqA},\eqref{eq:Dpqhsym},\eqref{eq:hstarsym}:

\begin{align}
	\Delta \sigma + D(p_{X_\text{fin}}\|p_{X_\text{ini}})%&= D(p\|\tilde{p})  \\
%	&= p_A(1) D(p\|\tilde{p}|A=1)\\
%	&\geq p_A(1) h^*_\text{sym}(d_{TV}(p,\tilde{p}|A=1))\\
	&\geq \frac{\langle A \rangle}{A_{\max}}  h^*_\text{sym}\left(d_{TV}(\frac{Ap}{\langle A \rangle},\frac{A\tilde{p}}{\langle A \rangle}) \right) \nonumber\\  
	&=\frac{\langle A \rangle}{A_{\max}}  h^*_\text{sym}\left(\frac{\langle A \,\, \text{sign}(p-\tilde{p}) \rangle}{\langle A \rangle} \right) \nonumber \\
	&=2 \frac{\langle A \,\, \text{sign}(p-\tilde{p}) \rangle}{A_{\max}}  \text{atanh} \, \frac{\langle A \,\, \text{sign}(p-\tilde{p}) \rangle}{\langle A \rangle}.
  \label{eq:crsigmahsymD}	 
\end{align}
%Here again, $\langle A \rangle$ is the probability that $X_\text{ini}\neq X_\text{fin}$.
%Here $f=A \,\, \text{sign}(p-\tilde{p})$ is a time-\emph{antisymmetric} observable built from $A$, i.e.  $f(\omega)=-f(\overline{\omega})$.. 
This is our first Thermo-Kinetic Relation  for constant-rate or time-symmetric-rate Markov chains, relating entropy production to nonnegative symmetric observables. Recall that $h^*_\text{sym}$ is the convex increasing function defined by \eqref{eq:hstarsym}. %This is one of the main result of this article.

Given \emph{any} observable $f: \Omega \to \mathbb{R}$ on trajectories, we can build the nonnegative time-symmetric observable $A(\omega)=\frac{|f(\omega)|+|f(\overline{\omega})|}{2}$.
Thus we get, for any trajectory observable $f: \Omega \to \mathbb{R}$:
\begin{align}
	\Delta \sigma + D(p_{X_\text{fin}}\|p_{X_\text{ini}})%&= D(p\|\tilde{p})  \\
	%	&= p_A(1) D(p\|\tilde{p}|A=1)\\
	%	&\geq p_A(1) h^*_\text{sym}(d_{TV}(p,\tilde{p}|A=1))\\  
	\geq \frac{\langle |f| \rangle}{|f|_{\max}}  h^*_\text{sym}\left(\frac{\langle \frac{|f|+|\overline{f}|}{2} \text{sign}(p-\tilde{p}) \rangle}{\langle |f| \rangle} \right).
	\label{eq:crsigmahanyf}	 
\end{align}
This implies a particularly simple form for time-\emph{antisymmetric} (also called \emph{current-like}, i.e. such that $f(\omega)=-f(\overline{\omega})$) observable $f$:

\begin{align}
	\Delta \sigma + D(p_{X_\text{fin}}\|p_{X_\text{ini}})
&\geq \frac{\langle |f| \rangle}{f_{\max}}  h^*_\text{sym}\left(\frac{|\langle f \rangle|}{\langle |f| \rangle} \right) \nonumber\\
&=  2\frac{|\langle f \rangle|}{f_{\max}}  \text{atanh}\frac{|\langle f \rangle|}{\langle |f| \rangle} .
	\label{eq:crsigmahantisym}	 
\end{align}
This is equivalent to the Thermo-Kinetic Relation \eqref{eq:crsigmahsymD} above, expressed for time-antisymmetric trajectory observables instead. This is the first main result of this article.

The r.h.s. is increasing in the $|\langle f \rangle|$ argument  and decreasing in the $\langle |f| \rangle$ argument %(from convexity of $h^*_{\text{sym}}$) 
in the $f_{\max}$ argument. From these observations we may infer new valid inequalities, e.g. by replacing $\langle |f| \rangle$ with an upper bound in the r.h.s.  For example the trivial bound $\langle |f| \rangle \leq f_{\max}$ leads to the simpler, but weaker, Thermo-Kinetic Relation:

   \begin{align}
   	\Delta \sigma + D(p_{X_\text{fin}}\|p_{X_\text{ini}})
   	\geq h^*_\text{sym}\left(\frac{|\langle f \rangle|}{f_{\max}} \right).
   	\label{eq:crsigmahantisymmax}	 
   \end{align}

%The Thermo-Kinetic Relation \eqref{eq:crsigmahantisym} is similar to the Thermodynamic Uncertainty Relation developed in CITE[Van Vu, Falasco, Potter, other?] in that it proposes a lower bound on entropy production on a constant or time-symmetric protocol from the statistics of an observable. 
%In fact it is shown in the Appendix that we can recover this Thermodynamic Uncertainty Relation (based on the mean and variance of the observable) from the present one. This means that if we seek to lower bound the  entropy production from the knowledge of an appropriate antisymmetric observable, then a better bound is obtained in general from the mean, mean absolute value and maximum than from mean and variance. Moreover the present version also works for non-stationary processes. 

A similar weakening of \eqref{eq:crsigmahantisym} leads to the following relation, still for anti-symmetric observables  $f$  and constant-rate or time-symmetric-rate Markov chains (details in the Appendix):
\begin{align}
\Delta \sigma   + D(p_{X_\text{fin}}\|p_{X_\text{ini}})  \geq 4\frac{ \langle f \rangle^2 }{\langle f^2 \rangle}\text{atanh}\left(\frac{ \langle f \rangle^2 / \langle f^2 \rangle}{ d_{TV}(p,\tilde{p})}\right).	\label{eq:ourTUR}
\end{align}

The $\text{atanh}$ term ranges from $0$ to infinity. 
This can be compared directly with the standard Thermodynamic Uncertainty Relation in the stationary case ($p_{X_\text{fin}}=p_{X_\text{ini}}$):

\begin{align}\label{eq:classicTUR}
	\Delta \sigma  \geq 2\frac{ \langle f \rangle^2 }{\text{Var} f },
\end{align}
with $\text{Var} f = \langle f \rangle^2 - \langle f^2 \rangle$.
This relation is tighest for an infinitesimal time interval, and increasingly looser for longer time interval \cite{falasco2020unifying}. In the infinitesimal time  limit $\tau_\text{fin}-\tau_\text{ini}=dt$, we have $\text{Var} f=\langle f^2 \rangle$ up to $\mathcal{O}(dt^2)$. Thus  \eqref{eq:ourTUR} and \eqref{eq:classicTUR} are easy to compare as they only differ by twice the $\text{atanh}$ factor.

We see in the Appendix that with the best possible choice of $f$ (with the maximum mean-to-standard-deviation ratio, or \emph{hyperaccurate current} in the terminology of \cite{busiello2019hyperaccurate}, making both \eqref{eq:ourTUR} and \eqref{eq:classicTUR} as tight as they can be), the novel Thermodynamic Uncertainty Relation  \eqref{eq:ourTUR},  a weakening of our general Thermo-Kinetic Relation  \eqref{eq:crsigmahantisym}, leads to a relation that can be weaker (when close to equilibrium, $p \approx \tilde{p}$) or tighter (when far from equilibrium, $p/\tilde{p} \to 0$ or $\infty$) than the classic \eqref{eq:classicTUR}. It has the added advantage that it is valid for nonstationary cases as well. 

In section V.C we will offer a numerical illustration of our Thermo-Kinetic Relation \eqref{eq:crsigmahantisym}  and the Thermodynamic Uncertainty Relation \eqref{eq:classicTUR} on a biophysical example---kinesin moving along a microtubule.

%It turns out, as shown in the Appendix, that this is precisely the Thermodynamics Uncertainty Relation for constant-rate or time-symmetric-rate Markov chains as proved in CITE(Van Vu, Falasco, Potter)  --- improved here in that it applies for non-stationary cases ($p_{X_\text{fin}}\neq p_{X_\text{ini}}$).
%This shows that the Thermo-Kinetic Relation above is stronger, in that it leads to tighter bounds on the entropy production than the Thermodynamic Uncertainty Relation, whenever the latter is not tight. 

\subsection{General time-varying Markov processes} Let us now turn to the general case of  time-varying rates over an arbitrary interval. We can decompose a time-varying Markov process into a succession of constant-rate Markov processes over infinitesimal intervals, in which at most one jump happens. Over each infinitesimal interval $dt$, we can thus use \eqref{eq:crsigmahantisym} (neglecting the $D(p_{X_\text{fin}}\|p_{X_\text{ini}})$ term, which scales as $dt^2$ over a $dt$ interval). Integrating  \eqref{eq:crsigmahantisym} on the total interval, and using convexity of $h^*_\text{sym}$, we find the Thermo-Kinetic Relation for general time-varying (overdamped) Markov chains, the second central result of this article: 

\begin{align} \label{eq:sigma-hsym}
	\Delta \sigma &\geq    \frac{\int |\langle f\rangle| dt}{f_{\max}}\,\,\,  h^*_\text{sym}\left(\frac{|\int \langle f  \rangle dt|}{ \int \langle |f| \rangle dt}\right)\\
	&=2\frac{|\int \langle f\rangle dt|}{f_{\max}}\,\,\,  \text{atanh} \frac{|\int \langle f  \rangle dt|}{ \int \langle |f| \rangle dt}.
\end{align}

%\begin{align} \label{eq:sigma-hsym}
%	\Delta \sigma \geq  \int  \frac{\langle A\rangle dt}{A_{\max}}\,\,\,  h^*_\text{sym}\left(\frac{\ell_{K_c}(p_{X_\text{ini}},p_{X_\text{fin}})}{ \int \langle A\rangle dt}\right)
%\end{align}

Notice that here $f$, an antisymmetric observable on infinitesimal-time trajectories, simply amounts to a weight $f(e)$ associated to every transition $e$ of the Markov chain, so that $f(e)=-f(\overline{e})$. Thus $f_{\max}$ is the maximum value of $f(e)$ over all transitions $e$. We can admit here observables that are themselves time-varying, i.e. the weight $f(e,t)$ assigned to a transition $e$ over the time interval $[t,t+dt]$ indeed depends on $t$. In this case, $f_{\max}$ is taken as the maximum (or supremum) over all possible transitions over all times. 

%Following the same loosening as in EQREF, we find the weaker bound:
%\begin{align} \label{eq:sigma-hsym}
%	\Delta \sigma  \geq     h^*_\text{sym}\left(\frac{|\int \langle f  \rangle dt|}{ f_{\max} \Delta t}\right)
%\end{align}

\section{Classical Speed Limits for general optimal transport costs, including the Total Variation distance} 
\label{sec:3}

The relations above are dependent on the particular protocol driving the initial distribution towards the final distribution. In particular they make sense even for stationary processes. It can be useful however to reduce the dependence of the formula above on the specifics of the particular process, in order to get an absolute lower bound on the entropy production cost required to drive the system from a given state distribution $p_\text{ini}$ to another state distribution $p_\text{fin}$. Strictly speaking, this approach is vain, in that in most cases one can always reach a state distribution from another at arbitrarily low entropy production cost, provided that the time-varying dynamics is sufficiently slow. Thus we can only hope to obtain lower bounds on entropy production relatedly to kinetic variables, in the (informal) form of: `If the driving is this fast, then the entropy production must be that high'. Such bounds are called (classical) \emph{Speed Limits} \cite{shi18,vo20,vanvu21,dechant2022minimum,tasnim2021thermodynamic}.

Measuring how `fast' a system is can be done in a number of ways, the relevance of which are context- and application-dependent, thus one cannot hope to obtain a `universal' speed limit theorem of interest but rather one speed limit for each situation of interest. A specific measure of `fastness' that has been used in most speed limits in the literature is the total expected \emph{activity}, i.e. the expected number of jumps of a discrete-state Markov chain over the total time interval: a fast process uses few jumps to reach the desired final distribution. While intuitive and relevant in number of circumstances, this measure has the disadvantage of being sensitive to the particular model chosen to capture the physical system. Indeed, a same physical system can often be modelled at different degrees of accuracy, with wildly different numbers of states. Models that are (exactly or approximately) equivalent from a kinetic and thermodynamic viewpoint may therefore exhibit wildly different activity, thus potentially lead to very loose speed limits. It seems more relevant in general to base bounds on physical observables, whose behaviour is not too dependent of an arbitrary modelling choice. The chief contribution of this section is precisely to exploit the Thermo-Kinetic relations derived above in order to offer a flexible family of Speed Limits based on statistics of physical observables, rather than the sole activity.

Consider that $A$, a time-symmetric nonnegative weight assigned to every trajectory $\omega$ in an arbitrary trajectory space $\Omega$, represents the `cost'  of $\omega$. For   each state $x \in X$ we assume we have a zero-cost trajectory $\omega_x$ going from $x$ to $x$ (e.g. the constant trajectory). 
%Then for any two states $x,x'$ we can define a cost $a(x,x')$ as the smallest possible cost $A(\omega)$ over all trajectories $\omega$ with $X_\text{ini}(\omega)=x$ and $X_\text{fin}(\omega)=x'$. Thus $a(x,x')=a(x',x)\geq 0$ and $a(x,x)=0$.

%$p^\text{Kant}$
 
  Given two probability distributions $p_\text{ini}$ and $p_\text{fin}$, we may now look for the probability distribution $p^*(\omega)$ on trajectories in $\Omega$ that drives the states from $p_\text{ini}$ to $p_\text{fin}$ with minimal cost $\langle A \rangle_{p^*}$. In other words we want to pick $p^*$ so that $p^*_{X_\text{ini}}=p_\text{ini}$ and $p^*_{X_\text{fin}}=p_\text{fin}$, while achieving the minimum possible value for $\langle A \rangle$. Finding such an optimal $p$ is essentially the standard \emph{optimal transport} problem, as introduced by Monge \cite{monge1781memoire} and by Kantorovich \cite{kantorovich2006translocation}. We call the optimal average cost $\langle A \rangle_{p}$ the \emph{Kantorovich cost} $K_A(p_\text{ini},p_\text{fin})$. Note that $K_A$ is not necessarily a proper distance on  the space on probability distributions on the state space $X$, although it is for many reasonable choices of $\Omega$ and $A$. 
  
 %  Note that $a$ is not necessarily a distance on $X$, for instance it may fail to satisfy the triangle inequality. However in case it is actually a proper distance on $X$, then $K_A$ is a proper distance on the space on probability distributions on the state space $X$.
  
 Clearly the  average cost given by any probability distribution $p$ on trajectories $\Omega$ is an upper bound on the Kantorovich cost between the initial and final state distributions $p_\text{ini}=p_{X_\text{ini}}$ and $p_\text{fin}=p_{X_\text{fin}}$:
 
 \begin{align}
 	\langle A \rangle_p \geq K_A(p_{X_\text{ini}},p_{X_\text{fin}}).
 \end{align}
 
 Nevertheless we can obtain a better bound on $K_A(p_{X_\text{ini}},p_{X_\text{fin}})$  by `improving' the probability distribution $p$. From $p$ we construct $p_\text{red}$,  a \emph{reduced}  probability distribution on trajectories, in the following way. For each $\omega$ from $x$ to $x' \neq x$ such that $p(\omega) >  p(\overline{\omega})$, set $p_\text{red}(\omega)=p(\omega) -  p(\overline{\omega})$, and $p_\text{red}(\overline{\omega})=0$. We also pick a zero-cost trajectory $\omega_x$ from $x$ to $x$ assign $p_\text{red}(\omega_x)=p(\omega_x) +  p(\overline{\omega})$. Similarly we  pick a zero-cost trajectory $\omega_{x'}$ from $x'$ to $x'$ and set $p_\text{red}(\omega_{x'})=p(\omega_{x'}) +  p(\overline{\omega})$. In other words we redirect part of the probability flow on self-loops in order to  `cancel' the flow along opposite directions. This keeps the probability distribution of initial and final states unperturbed, while decreasing the expected cost of trajectories. 
Thus we can write
 
 \begin{align}
 	\langle A \rangle_p \geq \langle A \rangle_{p_\text{red}}  \geq K_A(p_{X_\text{ini}},p_\text{fin}).
 \end{align}
 The middle quantity can be rewritten as
 \begin{align}
 	\langle A \rangle_{p_\text{red}}&=\frac{1}{2}\sum_\omega |Ap(\omega)-Ap(\overline{\omega})| \label{eq:pred}\\
 	&= \langle A \rangle_p d_{TV} \left (\frac{Ap}{\langle A \rangle_p},\frac{A\tilde{p}}{\langle A \rangle_p} \right).
 \end{align} 
  
  Thus, we can weaken \eqref{eq:crsigmahsymD} and obtain the following Speed Limit for constant-rate or time-symmetric-rate Markov chains:
   \begin{align} \label{eq:sigmaKAconstant}
   	  	\Delta \sigma + D(p_{X_\text{fin}}\|p_{X_\text{ini}}) \geq   \frac{\langle A \rangle_p}{A_{\max}} h^*_\text{sym}\left(\frac{K_A(p_{\text{ini}},p_{\text{fin}})}{\langle A \rangle_p}\right).
   \end{align}

 We now consider arbitrary time-varying Markov chains. A trajectory can be decomposed into constant-rate Markov chains over infinitesimal time intervals, where at most one transition happens. The cost $A$ on each interval is thus a cost assigned to each possible transition (as we request that constant paths have zero cost). This cost can possibly be  time-varying.   
 Integrating \eqref{eq:sigmaKAconstant} over the whole trajectory over the time interval:
  
 \begin{align} \label{eq:sigma-hsym-transport}
 	\Delta \sigma \geq    \frac{\int \langle A\rangle dt}{A_{\max}}\,\,\,  h^*_\text{sym}\left(\frac{\int dK_A )}{ \int \langle A\rangle dt}\right).
 \end{align}
where $\int dK_A$ is the integral of costs $K_A(p_{X_t},_{X_{t+dt}})$  over the whole trajectory from $p_{\text{ini}}$ to $p_{\text{fin}}$. Here $A_{\max}$ is the maximum value over all transistions and times, and $\int \langle A\rangle dt$ is the expected sum of costs of all transitions over the trajectory. 

The latter quantity can also be interpreted in the following way. Let us assign a  cost $\int A(\omega)$ of a trajectory $\omega$ over the whole interval $[\tau_\text{ini}, \tau_\text{fin}]$ defined as the integral of costs over infinitesimal transitions (i.e., the sum of cost of each transition along the trajectory, as we request that constant paths have zero cost). Then $\int \langle A\rangle dt$ is the mean value of this trajectory cost. As a result, the corresponding Kantorovich cost $K_{\int A}(p_\text{ini},p_\text{fin})$ is less than the integral of Kantorovich costs over infinitesimal intervals. 
Thus we may write
	\begin{align} \label{eq:sigma-hsym2}
		\Delta \sigma \geq    \frac{\int \langle A\rangle dt}{A_{\max}}\,\,\,  h^*_\text{sym}\left(\frac{K_{\int A}(p_{\text{ini}},p_{\text{fin}})}{ \int \langle A\rangle dt}\right).
	\end{align}
This Speed Limit has the advantage that $K_{\int A}(p_{\text{ini}},p_{\text{fin}})$ does not depend on the particular process driving  $p_{\text{ini}}$ to $p_{\text{fin}}$. The quantity $\int \langle A\rangle dt$ is the expected total (integrated) cost of the whole trajectory. %This is related  and is thus  indicative of the time taken by the protocol, as e.g. a slow protocol will integrate over a larger time interval. 

An important example of the above is the activity speed limit. Consider $A(\omega)=1$ for the extremities of the trajectory $\omega$ are distinct, and $A(\omega)=0$ if the extremities are identical. It is well known that the corresponding Kantorovich cost is precisely the Total Variation distance $d_{TV}$. Thus for any constant-rate or time-symmetric-rate Markov chain over an arbitrary Markov chain we find:

 \begin{align} \label{eq:constantfullacti}
	\Delta \sigma + D(p_{X_\text{fin}}\|p_{X_\text{ini}}) \geq   \langle A \rangle_p h^*_\text{sym}\left(\frac{d_{TV}(p_{\text{ini}},p_{\text{fin}})}{\langle A \rangle_p}\right),
\end{align}
where $\langle A \rangle_p$ is here the probability that $X_\text{fin} \neq X_\text{ini}$. For an arbitrary time-varying Markov chain we find:

\begin{align} \label{eq:cslactfull}
	\Delta \sigma  &\geq  \int \langle A \rangle_p dt \,\,\, h^*_\text{sym}\left(\frac{d_{TV}(p_{\text{ini}},p_{\text{fin}})}{\int \langle A \rangle_p dt}\right)\\
	&=2 d_{TV}(p_{\text{ini}},p_{\text{fin}}) \,\text{atanh}\,\frac{d_{TV}(p_{\text{ini}},p_{\text{fin}})}{\int \langle A \rangle_p dt},
\end{align}
where $\int \langle A \rangle_p  dt$ is the total expected activity, i.e. the expected number of transitions (jumps) along the total interval $[\tau_\text{ini}, \tau_\text{fin}]$. 
This number of transitions is the kinetic parameter indicative of the slowness of the dynamics. Fewer steps is thought of as a faster dynamics. This formula is obtained in \cite{dechant2022minimum}.
%, where it is also tight, i.e. can be achieved with a specific Markov protocol which occurs to involve no housekeeping entropy production ($\Delta \sigma_{NA}=\Delta \sigma$, conservative forces).

In comparison, the general formula \eqref{eq:sigma-hsym-transport} allows arbitrary (symmetric, nonnegative, time-varying) costs for each transition of the  symmetric weights. This allows to possibly model a physical cost of interest associated to the transition, and can mitigate the intrinsic dependence of activity on an arbitrary coarse graining level of the model at hand.

\section{Bounding non-adiabatic entropy production} 

\label{sec:4}

The total entropy $\Delta \sigma$, used in the speed limits above, decomposes into non-adiabatic (NA) entropy, and housekeeping entropy \eqref{eq:NAplusHH}.  Only the NA entropy is associated to the process of convergence to stationarity, while the housekeeping entropy may be non zero even at stationarity. In case of non-conservative forces, \eqref{eq:sigma-hsym} is thus typically far from tightness, especially at stationarity (where the r.h.s. side is zero and the l.h.s. is strictly positive) or around. 
The solution to this defect is to provide a bound using $\Delta \sigma_{NA}$ as the thermodynamic side of the relation \cite{shi18,vo20}. 

\subsection{Constant or time-symmetric rates}
 
Let us first focus on the case of constant-rate or time-symmetric-rate Markov processes, with corresponding stationary distribution $q$ on paths. Assume we have a nonnegative symmetric observable $A$ on the paths, i.e. a map $A:\Omega \to \mathbb{R}^+$ such that $A(\omega)=A(\overline{\omega})$, for any path $\omega$. Thus in particular we have for instance $q_{A} =\tilde{q}_{A}=\overline{q}_{A}$. On the other hand we know from stationarity that $q_{ X_\text{fin}}=\tilde{q}_{ X_\text{fin}}=\overline{q}_{X_\text{fin}}=q_{ X_\text{ini}}$. Now assume the stronger  that moreover $A$ is such that the equality holds for the joint observable $(A, X_\text{fin})$, i.e. $q_{A X_\text{fin}}=\tilde{q}_{A X_\text{fin}}=\overline{q}_{A X_\text{fin}}$. In other words, even though the stationary process might not satisfy detailed balance ($q\neq \overline{q}$), we have `apparent' detailed balance from the joint knowledge of $A$ and $X_{\text{fin}}$. This is the case for instance for the `activity' observable $A: \Omega \to \{0,1\}$, mapping a path $\omega$ to $1$ if $X_\text{ini} \neq X_\text{fin}$ (the path starts and ends in different states) and to $0$ if $X_\text{ini} = X_\text{fin}$. Indeed the probability $q(A=1, X_\text{fin}=x)$ records the probability flow entering $x$ over the trajectory (from any initial state than $x$). On the other hand $\tilde{q}(A=1, X_\text{fin}=x)$ is the probability flow leaving the initial state $x$. From stationarity, these quantities must be equal. Other observables than activity may be eligible, depending on the specificities of the system at hand, as we see later in the case of electronic memories. Let us restate the property that such an observable must satisfy. For each possible value of $A=z$ of such an observable, the corresponding level set $\{\omega \in \Omega: A(\omega)=z\}$ has probability flow conservation at each state $x$ at stationarity. In other words, the global probability flow conservation at stationarity splits into several conservation laws on several sets of paths. 

% FOR EXAMPLES IN MEMORIES, ELECTRIC CURRENT? VOLTAGE?

%To find a lower bound on NA entropy production,  we only keep from the path the information of the final state and whether the initial state differs from the final state. Consider indeed the `activity' observable $A: \Omega \to \{0,1\}$, mapping a path $\omega$ to $1$ if $X_\text{ini} \neq X_\text{fin}$ (the path starts and ends in different states) and to $0$ if $X_\text{ini} = X_\text{fin}$. Then $p_{AX_\text{fin}}$ and $q_{AX_\text{fin}}$ are distributions on $\{0,1\} \times \mathcal{X}$ and we can write:

To find a lower bound on NA entropy production, we write the following:
\begin{align}
	D(p\|q| X_\text{fin}) &\geq D(p_{AX_\text{fin}}\|q_{AX_\text{fin}}|X_\text{fin}) \\
	&=D(p_{AX_\text{fin}}\|\overline{q}_{AX_\text{fin}}| X_\text{fin})\\
	&=D(p_{AX_\text{fin}}\|\overline{p}_{AX_\text{fin}}| X_\text{fin})\\
	&=D(p_{AX_\text{fin}}\|\tilde{p}_{AX_\text{fin}}| X_\text{fin}).\label{eq:boundXfin}
\end{align}
We used here the fact that $q_{AX_\text{fin}}=\overline{q}_{AX_\text{fin}}$ and that $\overline{q}$, $\overline{p}$ and $\tilde{p}$ coincide when knowing the value of $X_\text{fin}$ (initial state of the reversed path). 
We find by repeated use of the Chain Rule, \eqref{eq:boundXfin},\eqref{eq:DpqA}, and \eqref{eq:DpqhdTV} 
\begin{align}
	\Delta \sigma_{NA}&+D(p_{X_\text{fin}}\|p_{X_\text{ini}})\\
	&\geq D(p_{AX_\text{fin}}\|\tilde{p}_{AX_\text{fin}}| X_\text{fin})+D(p_{X_\text{fin}}\|p_{X_\text{ini}})\\
	&=D(p_{AX_\text{fin}}\|\tilde{p}_{AX_\text{fin}}| X_\text{fin})+D(p_{X_\text{fin}}\|\tilde{p}_{X_\text{fin}})\\
	&=D(p_{AX_\text{fin}}\|\tilde{p}_{AX_\text{fin}})\\ 
	&\geq \frac{\langle A \rangle_p}{A_\text{max}} D(\frac{Ap_{AX_\text{fin}}}{\langle A \rangle_p}\|\frac{A\tilde{p}_{AX_\text{fin}}}{\langle A \rangle_p})\\
	&=\frac{\langle A \rangle_p}{A_\text{max}} D(\frac{Ap_{AX_\text{fin}}}{\langle A \rangle_p}\|\frac{Ap_{AX_\text{ini}}}{\langle A \rangle_p})\\
	&=\frac{\langle A \rangle_p}{A_\text{max}} h(d_{TV}(\frac{Ap_{AX_\text{fin}}}{\langle A \rangle_p},\frac{Ap_{AX_\text{ini}}}{\langle A \rangle_p}))	\\
	&=\frac{\langle A \rangle_p}{A_\text{max}} h\left(\frac{\langle A\,\, \text{sign}(p_{AX_\text{fin}}-p_{AX_\text{ini}})\rangle}{\langle A \rangle_p} \right). \label{eq:TKRNAvsA}
\end{align}
%In summary, for any constant-rate or time-symmetric-rate Markov process, over any interval of time, and any nonnegative symmetric observable with the property above, we have
%\begin{align} \label{eq:cr-HS-TVA}
%	\Delta \sigma_{NA} + D(p_{X_\text{fin}}\|p_{X_\text{ini}}) \geq \langle A \rangle h\left(\frac{d_{TV}(p_{X_\text{fin}},p_{X_\text{ini}})}{\langle A \rangle}\right).
%\end{align}

This is our main Thermo-Kinetic Relation, for non-adiabatic entropy production, for constant-rate or time-symmetric-rate Markov chains. Remember that $h$ can taken as Pinsker's or Gilardoni's bound \eqref{eq:gilard}, or the optimal (non-explicit) optimal bound $h^*$, as pictured in Fig.~\eqref{fig:bounds}.
We can reformulate it in terms of time-antisymmetric observables, similarly to \eqref{eq:crsigmahantisym}. Let  $f$ be an anti-symmetric observable on trajectories, such that at every state $x$ and for every possible value $f=z$, the probability flow carried by trajectories of value $z$ entering $x$ equals, for the stationary flow $q$, the probability flow carried by trajectories of value $-z$ out of $x$. Then  we find:
\begin{align} \label{eq:constsigmaNAantisym}
	\Delta \sigma_{NA}+D(p_{X_\text{fin}}\|p_{X_\text{ini}})\geq \frac{\langle |f| \rangle}{f_\text{max}} h\left(\frac{|\langle f \rangle|}{\langle |f| \rangle} \right).
\end{align}

We can also bring \eqref{eq:TKRNAvsA} to an optimal transport interpretation:
\begin{align}\label{eq:constsigmaNAtransport}
	\Delta \sigma_{NA}&+D(p_{X_\text{fin}}\|p_{X_\text{ini}}) \nonumber\\
	&\geq\frac{\langle A \rangle_p}{A_\text{max}} h\left(\frac{K_A(p_{X_\text{ini}},p_{X_\text{fin}})}{\langle A \rangle_p} \right).
\end{align}
In particular for $A$ the activity observable, i.e. when $\langle A \rangle_p$ is the probability that $X_\text{ini} \neq X_\text{fin}$, we find
\begin{align}
	\Delta \sigma_{NA}&+D(p_{X_\text{fin}}\|p_{X_\text{ini}}) \nonumber\\
	&\geq\frac{\langle A \rangle_p}{A_\text{max}} h\left(\frac{d_{TV}(p_{X_\text{ini}},p_{X_\text{fin}})}{\langle A \rangle_p} \right).
\end{align}

\subsection{General time-varying Markov chains: the classical Speed Limit for non-adiabatic entropy production} Let us now turn to the general case of  time-varying rates over an arbitrary interval. We can decompose a time-varying Markov process into a succession of constant-rate Markov processes over infinitesimal intervals. Over each infinitesimal interval $dt$, we can thus use  \eqref{eq:constsigmaNAtransport} (neglecting the $D(p_{X_\text{fin}}\|p_{X_\text{ini}})$ term). Integrating  \eqref{eq:constsigmaNAtransport} on the total interval, and using convexity of $h$, we find 
%
%\begin{align}\label{eq:CSL-infinitesimal}
%	d_{TV}(p_{AX_\text{fin}}\|\overline{p}_{AX_\text{fin}}|A=1)=\frac{d_{TV}(p_{X_\text{ini}},p_{X_\text{fin}})}{\int \langle A(t) \rangle dt}
%\end{align}
%
\begin{align} \label{eq:sigmaHS-h}
	\Delta \sigma_{NA} &\geq \frac{ \int \langle A\rangle dt}{A_{\max}} \,\,\,\, h \left( \frac{\int dK_A}{ \int \langle A\rangle dt} \right) \nonumber\\&\geq  \frac{ \int \langle A\rangle dt}{A_{\max}} \,\,\,\, h \left( \frac{ K_{\int A}(p_{\text{ini}},p_{\text{fin}})}{ \int \langle A\rangle dt} \right),	
\end{align}
where $\int dK_A$  is the integrated Kantorovitch cost along the trajectory as in \eqref{eq:sigma-hsym}, and can be weakened to the trajectory-independent quantity $K_{\int A}(p_{\text{ini}},p_{\text{fin}})$ as in \eqref{eq:sigma-hsym2}.
This Thermo-Kinetic Relation is the third central result of this article, and can also be called a classical Speed Limit. 

In the case of the  activity  observable, we obtain:
\begin{align} \label{eq:sigmaHS-h-TV}
	\Delta \sigma_{NA} \geq  \int \langle A\rangle dt \,\,\,\, h \left( \frac{d_{TV}(p_{X_\text{fin}},p_{X_\text{ini}})}{ \int \langle A\rangle dt} \right),	
\end{align}
where $\int \langle A\rangle dt$ the total expected activity, i.e. the expected number of jumps in the whole time interval $\Delta t$.
Recall that here $h$ can be taken as $h(x)=2x^2$ (Pinsker's bound),
leading to  
\begin{align}
	\Delta \sigma_{NA} \geq  2\frac{d^2_{TV}(p_\text{ini},p_\text{fin} )}{ \int \langle A\rangle dt}
\end{align}
This is essentially the main result of \cite{vo20}, itself a refinement of  \cite{shi18}. The bound is substantially tighter, using e.g. Gilardoni's bound \eqref{eq:gilard}.
%From Gilardoni's bound ($h(x)=\ln\frac{1}{1-x} - (1-x)\ln(1+x)$) we obtain the following novel Classical Speed Limit for activity, which we recommend as being both accurate and convenient:
%\begin{align} \label{eq:CSLGilardoni}
%	\Delta  \sigma_{NA} &\geq \int \langle A\rangle dt \ln \frac{1}{1-\frac{d_{TV}}{ \int \langle A\rangle dt}}  \\&  - \left( \int \langle A\rangle dt-\ell_{TV} \right) \ln \left(1+\frac{d_{TV}}{ \int \langle A\rangle dt}\right)\nonumber
%\end{align}
Even slightly better is the bound obtained with $h=h^*$, although it has no simple explicit expression. All these bounds coincide in the limit of long times (high activity, slow driving). In the limit of low times or low activity (fast driving),  Gilardoni's classical Speed Limit is unboundedly better than Pinsker's. Indeed it  correctly predicts that the activity (expected number of jumps) to follow the trajectory must satisfy:
\begin{align}
	\int \langle A\rangle dt >  d_{TV}(p_\text{ini},p_\text{fin} ),
\end{align}
with an infinite NA entropy production at the limit case $\langle A \rangle =  d_{TV}$. In constrast, Pinsker's version only provides a finite lower bound. 

Finally, we also have a time-varying version of the Thermo-Kinetic Relation bound   \eqref{eq:constsigmaNAantisym} for anti-symmetric observable $f$ on transitions satisfying at all times the flow conservation condition:

\begin{align}
	\Delta \sigma_{NA}\geq \frac{\int \langle |f| \rangle dt}{f_\text{max}} h\left(\frac{|\langle f \rangle|}{\int \langle |f| \rangle dt} \right).
\end{align}

%\begin{align}
%\int \langle  A  \rangle dt=(\tau_\text{fin}-\tau_\text{ini}) \langle A_{\text{time-av}} \rangle
%\end{align}
%where $A_{\text{time-av}}$ is the time averaged expected activity along the trajectory. Thus if we constrain the expected number of jumps per time unit,  $\int \langle  A  \rangle dt$ serves as a measure of duration.

%From Vajda's bound ($h(x)=\ln\frac{1+x}{1-x} - 2\frac{x}{1+x}$)  we find:

%\begin{align}
%\Delta \sigma_{NA} \geq  \int \langle A\rangle dt \ln \frac{1+\frac{\ell_{TV}}{ \int \langle A\rangle dt}}{1-\frac{\ell_{TV}}{ \int \langle A\rangle dt}}  - 2\frac{  \ell_{TV}}{1+\frac{\ell_{TV}}{ \int \langle A\rangle dt}}
%\end{align}

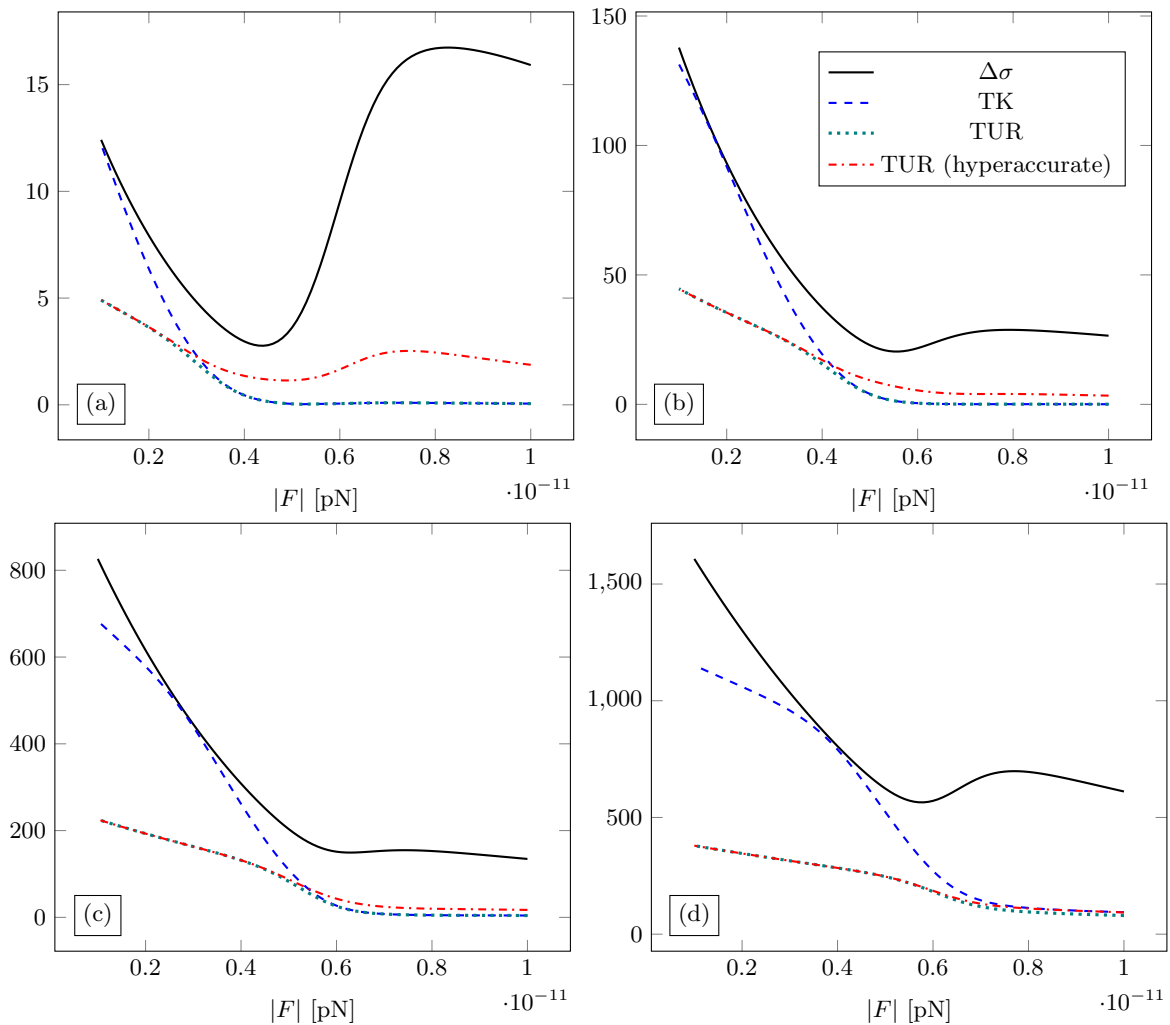
\begin{figure*}[t]
\center
\begin{tikzpicture}
\begin{axis}[xlabel={$|F|$ [pN]},
    legend style={at={(0.05,0.75)},anchor=west}]
    \node[draw] at (0, 0)   (a) {(a)};
     \addplot[mark=none,thick,solid,black] coordinates {
     (1.*10^-11,15.9161)(9.9*10^-12,15.9885)(9.8*10^-12,16.059)(9.7*10^-12,16.1275)(9.6*10^-12,16.1939)(9.5*10^-12,16.258)(9.4*10^-12,16.3196)(9.3*10^-12,16.3785)(9.2*10^-12,16.4343)(9.1*10^-12,16.4869)(9.*10^-12,16.5358)(8.9*10^-12,16.5808)(8.8*10^-12,16.6212)(8.7*10^-12,16.6567)(8.6*10^-12,16.6865)(8.5*10^-12,16.7099)(8.4*10^-12,16.7262)(8.3*10^-12,16.7342)(8.2*10^-12,16.7329)(8.1*10^-12,16.7209)(8.*10^-12,16.6966)(7.9*10^-12,16.6583)(7.8*10^-12,16.6037)(7.7*10^-12,16.5306)(7.6*10^-12,16.4361)(7.5*10^-12,16.3173)(7.4*10^-12,16.1707)(7.3*10^-12,15.9927)(7.2*10^-12,15.7792)(7.1*10^-12,15.526)(7.*10^-12,15.2292)(6.9*10^-12,14.8846)(6.8*10^-12,14.4887)(6.7*10^-12,14.039)(6.6*10^-12,13.534)(6.5*10^-12,12.9738)(6.4*10^-12,12.3608)(6.3*10^-12,11.6996)(6.2*10^-12,10.9974)(6.1*10^-12,10.2639)(6.*10^-12,9.51097)(5.9*10^-12,8.75211)(5.8*10^-12,8.00154)(5.7*10^-12,7.27337)(5.6*10^-12,6.58065)(5.5*10^-12,5.93456)(5.4*10^-12,5.34386)(5.3*10^-12,4.81466)(5.2*10^-12,4.35035)(5.1*10^-12,3.95188)(5.*10^-12,3.61813)(4.9*10^-12,3.34639)(4.8*10^-12,3.13283)(4.7*10^-12,2.97294)(4.6*10^-12,2.86188)(4.5*10^-12,2.79481)(4.4*10^-12,2.76705)(4.3*10^-12,2.77427)(4.2*10^-12,2.81253)(4.1*10^-12,2.87834)(4.*10^-12,2.96865)(3.9*10^-12,3.08084)(3.8*10^-12,3.21271)(3.7*10^-12,3.3624)(3.6*10^-12,3.52839)(3.5*10^-12,3.70944)(3.4*10^-12,3.90457)(3.3*10^-12,4.11301)(3.2*10^-12,4.33417)(3.1*10^-12,4.56761)(3.*10^-12,4.81305)(2.9*10^-12,5.0703)(2.8*10^-12,5.33928)(2.7*10^-12,5.61997)(2.6*10^-12,5.91246)(2.5*10^-12,6.21685)(2.4*10^-12,6.53334)(2.3*10^-12,6.86213)(2.2*10^-12,7.2035)(2.1*10^-12,7.55772)(2.*10^-12,7.92514)(1.9*10^-12,8.30609)(1.8*10^-12,8.70095)(1.7*10^-12,9.11012)(1.6*10^-12,9.53401)(1.5*10^-12,9.97307)(1.4*10^-12,10.4277)(1.3*10^-12,10.8985)(1.2*10^-12,11.3859)(1.1*10^-12,11.8903)(1.*10^-12,12.4124)     };
        %    \addlegendentry{$\Delta \sigma$};
       \addplot[mark=none,thick,dashed,blue] coordinates {(1.*10^-11,0.0564263)(9.9*10^-12,0.0577343)(9.8*10^-12,0.0590708)(9.7*10^-12,0.0604362)(9.6*10^-12,0.0618306)(9.5*10^-12,0.0632541)(9.4*10^-12,0.0647065)(9.3*10^-12,0.0661874)(9.2*10^-12,0.0676965)(9.1*10^-12,0.0692328)(9.*10^-12,0.0707952)(8.9*10^-12,0.0723819)(8.8*10^-12,0.0739909)(8.7*10^-12,0.0756193)(8.6*10^-12,0.0772633)(8.5*10^-12,0.0789183)(8.4*10^-12,0.0805782)(8.3*10^-12,0.0822357)(8.2*10^-12,0.0838813)(8.1*10^-12,0.0855037)(8.*10^-12,0.0870887)(7.9*10^-12,0.0886189)(7.8*10^-12,0.0900735)(7.7*10^-12,0.091427)(7.6*10^-12,0.0926493)(7.5*10^-12,0.0937046)(7.4*10^-12,0.0945506)(7.3*10^-12,0.0951387)(7.2*10^-12,0.0954133)(7.1*10^-12,0.0953119)(7.*10^-12,0.0947661)(6.9*10^-12,0.0937035)(6.8*10^-12,0.0920495)(6.7*10^-12,0.0897318)(6.6*10^-12,0.0866855)(6.5*10^-12,0.0828595)(6.4*10^-12,0.0782239)(6.3*10^-12,0.0727781)(6.2*10^-12,0.0665585)(6.1*10^-12,0.0614579)(6.*10^-12,0.058239)(5.9*10^-12,0.0547359)(5.8*10^-12,0.051018)(5.7*10^-12,0.0471632)(5.6*10^-12,0.0432525)(5.5*10^-12,0.0393651)(5.4*10^-12,0.035573)(5.3*10^-12,0.0319374)(5.2*10^-12,0.0285066)(5.1*10^-12,0.0332017)(5.*10^-12,0.0438076)(4.9*10^-12,0.0572071)(4.8*10^-12,0.0740231)(4.7*10^-12,0.0949954)(4.6*10^-12,0.120995)(4.5*10^-12,0.153033)(4.4*10^-12,0.19227)(4.3*10^-12,0.240011)(4.2*10^-12,0.297701)(4.1*10^-12,0.366896)(4.*10^-12,0.44923)(3.9*10^-12,0.546371)(3.8*10^-12,0.659959)(3.7*10^-12,0.791546)(3.6*10^-12,0.942535)(3.5*10^-12,1.11413)(3.4*10^-12,1.3073)(3.3*10^-12,1.52277)(3.2*10^-12,1.76102)(3.1*10^-12,2.02231)(3.*10^-12,2.30669)(2.9*10^-12,2.61409)(2.8*10^-12,2.94433)(2.7*10^-12,3.29713)(2.6*10^-12,3.6722)(2.5*10^-12,4.06919)(2.4*10^-12,4.48774)(2.3*10^-12,4.92746)(2.2*10^-12,5.38793)(2.1*10^-12,5.86866)(2.*10^-12,6.36907)(1.9*10^-12,6.88847)(1.8*10^-12,7.42601)(1.7*10^-12,7.98065)(1.6*10^-12,8.55112)(1.5*10^-12,9.13588)(1.4*10^-12,9.73311)(1.3*10^-12,10.3407)(1.2*10^-12,10.9563)(1.1*10^-12,11.5772)(1.*10^-12,12.2008)    
         };
    %   \addlegendentry{TK};
         \addplot[mark=none,very thick,dotted,teal] coordinates {(1.*10^-11,0.0564211)(9.9*10^-12,0.0577288)(9.8*10^-12,0.0590652)(9.7*10^-12,0.0604304)(9.6*10^-12,0.0618245)(9.5*10^-12,0.0632478)(9.4*10^-12,0.0646999)(9.3*10^-12,0.0661806)(9.2*10^-12,0.0676894)(9.1*10^-12,0.0692255)(9.*10^-12,0.0707875)(8.9*10^-12,0.072374)(8.8*10^-12,0.0739827)(8.7*10^-12,0.0756108)(8.6*10^-12,0.0772545)(8.5*10^-12,0.0789091)(8.4*10^-12,0.0805687)(8.3*10^-12,0.0822259)(8.2*10^-12,0.0838712)(8.1*10^-12,0.0854932)(8.*10^-12,0.0870778)(7.9*10^-12,0.0886077)(7.8*10^-12,0.0900619)(7.7*10^-12,0.0914152)(7.6*10^-12,0.0926372)(7.5*10^-12,0.0936922)(7.4*10^-12,0.094538)(7.3*10^-12,0.0951259)(7.2*10^-12,0.0954003)(7.1*10^-12,0.0952988)(7.*10^-12,0.0947531)(6.9*10^-12,0.0936905)(6.8*10^-12,0.0920367)(6.7*10^-12,0.0897194)(6.6*10^-12,0.0866737)(6.5*10^-12,0.0828483)(6.4*10^-12,0.0782135)(6.3*10^-12,0.0727687)(6.2*10^-12,0.0665501)(6.1*10^-12,0.0614503)(6.*10^-12,0.0582317)(5.9*10^-12,0.054729)(5.8*10^-12,0.0510115)(5.7*10^-12,0.047157)(5.6*10^-12,0.0432468)(5.5*10^-12,0.0393598)(5.4*10^-12,0.0355681)(5.3*10^-12,0.0319329)(5.2*10^-12,0.0285026)(5.1*10^-12,0.0331955)(5.*10^-12,0.0437952)(4.9*10^-12,0.0571827)(4.8*10^-12,0.0739759)(4.7*10^-12,0.0949055)(4.6*10^-12,0.120826)(4.5*10^-12,0.15272)(4.4*10^-12,0.191697)(4.3*10^-12,0.23898)(4.2*10^-12,0.295872)(4.1*10^-12,0.363703)(4.*10^-12,0.443757)(3.9*10^-12,0.537162)(3.8*10^-12,0.644776)(3.7*10^-12,0.767048)(3.6*10^-12,0.903901)(3.5*10^-12,1.05464)(3.4*10^-12,1.21793)(3.3*10^-12,1.39182)(3.2*10^-12,1.57388)(3.1*10^-12,1.76137)(3.*10^-12,1.95151)(2.9*10^-12,2.14162)(2.8*10^-12,2.32938)(2.7*10^-12,2.51293)(2.6*10^-12,2.69093)(2.5*10^-12,2.86254)(2.4*10^-12,3.02742)(2.3*10^-12,3.1856)(2.2*10^-12,3.33741)(2.1*10^-12,3.48339)(2.*10^-12,3.62423)(1.9*10^-12,3.76067)(1.8*10^-12,3.89348)(1.7*10^-12,4.02339)(1.6*10^-12,4.15113)(1.5*10^-12,4.27736)(1.4*10^-12,4.40267)(1.3*10^-12,4.52762)(1.2*10^-12,4.65269)(1.1*10^-12,4.77831)(1.*10^-12,4.90488)
         };
%  \addlegendentry{TUR};
         \addplot[mark=none,thick,dashdotted,red] coordinates {(1.*10^-11,1.87425)(9.9*10^-12,1.90195)(9.8*10^-12,1.93001)(9.7*10^-12,1.95843)(9.6*10^-12,1.98719)(9.5*10^-12,2.01627)(9.4*10^-12,2.04566)(9.3*10^-12,2.07532)(9.2*10^-12,2.10523)(9.1*10^-12,2.13534)(9.*10^-12,2.16559)(8.9*10^-12,2.19593)(8.8*10^-12,2.22626)(8.7*10^-12,2.25648)(8.6*10^-12,2.28648)(8.5*10^-12,2.31608)(8.4*10^-12,2.34512)(8.3*10^-12,2.37335)(8.2*10^-12,2.40052)(8.1*10^-12,2.42628)(8.*10^-12,2.45025)(7.9*10^-12,2.47197)(7.8*10^-12,2.4909)(7.7*10^-12,2.50641)(7.6*10^-12,2.5178)(7.5*10^-12,2.52427)(7.4*10^-12,2.52494)(7.3*10^-12,2.51887)(7.2*10^-12,2.5051)(7.1*10^-12,2.4827)(7.*10^-12,2.45081)(6.9*10^-12,2.40877)(6.8*10^-12,2.35617)(6.7*10^-12,2.29299)(6.6*10^-12,2.2197)(6.5*10^-12,2.1373)(6.4*10^-12,2.04736)(6.3*10^-12,1.95196)(6.2*10^-12,1.85358)(6.1*10^-12,1.75492)(6.*10^-12,1.65869)(5.9*10^-12,1.56738)(5.8*10^-12,1.48307)(5.7*10^-12,1.40734)(5.6*10^-12,1.34119)(5.5*10^-12,1.28507)(5.4*10^-12,1.23898)(5.3*10^-12,1.20259)(5.2*10^-12,1.1753)(5.1*10^-12,1.15641)(5.*10^-12,1.1452)(4.9*10^-12,1.14097)(4.8*10^-12,1.14309)(4.7*10^-12,1.15107)(4.6*10^-12,1.16453)(4.5*10^-12,1.18323)(4.4*10^-12,1.20706)(4.3*10^-12,1.23607)(4.2*10^-12,1.27041)(4.1*10^-12,1.31039)(4.*10^-12,1.3564)(3.9*10^-12,1.40896)(3.8*10^-12,1.46863)(3.7*10^-12,1.53599)(3.6*10^-12,1.61162)(3.5*10^-12,1.69598)(3.4*10^-12,1.78938)(3.3*10^-12,1.89187)(3.2*10^-12,2.00323)(3.1*10^-12,2.12287)(3.*10^-12,2.24989)(2.9*10^-12,2.38308)(2.8*10^-12,2.52102)(2.7*10^-12,2.66221)(2.6*10^-12,2.80515)(2.5*10^-12,2.94848)(2.4*10^-12,3.09106)(2.3*10^-12,3.23203)(2.2*10^-12,3.37082)(2.1*10^-12,3.50712)(2.*10^-12,3.64087)(1.9*10^-12,3.77219)(1.8*10^-12,3.90134)(1.7*10^-12,4.02869)(1.6*10^-12,4.15465)(1.5*10^-12,4.27965)(1.4*10^-12,4.40413)(1.3*10^-12,4.52852)(1.2*10^-12,4.65323)(1.1*10^-12,4.77862)(1.*10^-12,4.90504)  };
% \addlegendentry{TUR ($f_\text{hyp}$)};
\end{axis}
\end{tikzpicture}
%%%
\begin{tikzpicture}
\begin{axis}[xlabel={$|F|$ [pN]},
    legend style={at={(0.95,0.75)},anchor=east}]
     \node[draw] at (0, 0)   (a) {(b)};
     \addplot[mark=none,thick,solid,black] coordinates {
     (1.*10^-11,26.5131)(9.9*10^-12,26.6653)(9.8*10^-12,26.816)(9.7*10^-12,26.9649)(9.6*10^-12,27.1118)(9.5*10^-12,27.2564)(9.4*10^-12,27.3986)(9.3*10^-12,27.5378)(9.2*10^-12,27.6738)(9.1*10^-12,27.8061)(9.*10^-12,27.934)(8.9*10^-12,28.0571)(8.8*10^-12,28.1745)(8.7*10^-12,28.2854)(8.6*10^-12,28.3888)(8.5*10^-12,28.4836)(8.4*10^-12,28.5684)(8.3*10^-12,28.6417)(8.2*10^-12,28.7017)(8.1*10^-12,28.7464)(8.*10^-12,28.7733)(7.9*10^-12,28.7798)(7.8*10^-12,28.7629)(7.7*10^-12,28.7194)(7.6*10^-12,28.6454)(7.5*10^-12,28.5371)(7.4*10^-12,28.3903)(7.3*10^-12,28.2007)(7.2*10^-12,27.964)(7.1*10^-12,27.6765)(7.*10^-12,27.335)(6.9*10^-12,26.9373)(6.8*10^-12,26.4831)(6.7*10^-12,25.974)(6.6*10^-12,25.4146)(6.5*10^-12,24.8127)(6.4*10^-12,24.1799)(6.3*10^-12,23.5317)(6.2*10^-12,22.8872)(6.1*10^-12,22.2686)(6.*10^-12,21.6999)(5.9*10^-12,21.2061)(5.8*10^-12,20.811)(5.7*10^-12,20.536)(5.6*10^-12,20.3987)(5.5*10^-12,20.4121)(5.4*10^-12,20.5845)(5.3*10^-12,20.919)(5.2*10^-12,21.4147)(5.1*10^-12,22.0672)(5.*10^-12,22.8697)(4.9*10^-12,23.8139)(4.8*10^-12,24.8906)(4.7*10^-12,26.0907)(4.6*10^-12,27.4057)(4.5*10^-12,28.8278)(4.4*10^-12,30.35)(4.3*10^-12,31.9667)(4.2*10^-12,33.6731)(4.1*10^-12,35.4658)(4.*10^-12,37.3419)(3.9*10^-12,39.2998)(3.8*10^-12,41.3383)(3.7*10^-12,43.457)(3.6*10^-12,45.656)(3.5*10^-12,47.9361)(3.4*10^-12,50.2982)(3.3*10^-12,52.7437)(3.2*10^-12,55.2742)(3.1*10^-12,57.8917)(3.*10^-12,60.5982)(2.9*10^-12,63.3961)(2.8*10^-12,66.2878)(2.7*10^-12,69.2759)(2.6*10^-12,72.3631)(2.5*10^-12,75.5522)(2.4*10^-12,78.846)(2.3*10^-12,82.2477)(2.2*10^-12,85.7603)(2.1*10^-12,89.387)(2.*10^-12,93.131)(1.9*10^-12,96.9957)(1.8*10^-12,100.984)(1.7*10^-12,105.101)(1.6*10^-12,109.348)(1.5*10^-12,113.731)(1.4*10^-12,118.252)(1.3*10^-12,122.915)(1.2*10^-12,127.725)(1.1*10^-12,132.685)(1.*10^-12,137.8) };
            \addlegendentry{$\Delta \sigma$};
         \addplot[mark=none,thick,dashed,blue] coordinates {(1.*10^-11,0.0563434)(9.9*10^-12,0.0576313)(9.8*10^-12,0.058943)(9.7*10^-12,0.0602776)(9.6*10^-12,0.0616337)(9.5*10^-12,0.0630097)(9.4*10^-12,0.0644032)(9.3*10^-12,0.0658112)(9.2*10^-12,0.0672297)(9.1*10^-12,0.0686538)(9.*10^-12,0.0700771)(8.9*10^-12,0.0714916)(8.8*10^-12,0.0728873)(8.7*10^-12,0.0742517)(8.6*10^-12,0.0755693)(8.5*10^-12,0.0768209)(8.4*10^-12,0.0779831)(8.3*10^-12,0.0790269)(8.2*10^-12,0.0799172)(8.1*10^-12,0.0806115)(8.*10^-12,0.0810587)(7.9*10^-12,0.0811978)(7.8*10^-12,0.0809571)(7.7*10^-12,0.0802531)(7.6*10^-12,0.0789906)(7.5*10^-12,0.077063)(7.4*10^-12,0.0743557)(7.3*10^-12,0.0720651)(7.2*10^-12,0.0727774)(7.1*10^-12,0.0760343)(7.*10^-12,0.0836259)(6.9*10^-12,0.093209)(6.8*10^-12,0.105362)(6.7*10^-12,0.12084)(6.6*10^-12,0.140626)(6.5*10^-12,0.165994)(6.4*10^-12,0.198585)(6.3*10^-12,0.240501)(6.2*10^-12,0.294396)(6.1*10^-12,0.363595)(6.*10^-12,0.452199)(5.9*10^-12,0.565199)(5.8*10^-12,0.708576)(5.7*10^-12,0.889376)(5.6*10^-12,1.11576)(5.5*10^-12,1.39702)(5.4*10^-12,1.74348)(5.3*10^-12,2.16641)(5.2*10^-12,2.67774)(5.1*10^-12,3.28981)(5.*10^-12,4.01491)(4.9*10^-12,4.86485)(4.8*10^-12,5.85048)(4.7*10^-12,6.98128)(4.6*10^-12,8.26497)(4.5*10^-12,9.7074)(4.4*10^-12,11.3124)(4.3*10^-12,13.082)(4.2*10^-12,15.0166)(4.1*10^-12,17.1153)(4.*10^-12,19.376)(3.9*10^-12,21.7963)(3.8*10^-12,24.3731)(3.7*10^-12,27.1031)(3.6*10^-12,29.9831)(3.5*10^-12,33.0096)(3.4*10^-12,36.1791)(3.3*10^-12,39.4878)(3.2*10^-12,42.9317)(3.1*10^-12,46.506)(3.*10^-12,50.2052)(2.9*10^-12,54.0228)(2.8*10^-12,57.9508)(2.7*10^-12,61.9796)(2.6*10^-12,66.0979)(2.5*10^-12,70.2923)(2.4*10^-12,74.5475)(2.3*10^-12,78.8461)(2.2*10^-12,83.1696)(2.1*10^-12,87.4981)(2.*10^-12,91.8119)(1.9*10^-12,96.0919)(1.8*10^-12,100.321)(1.7*10^-12,104.484)(1.6*10^-12,108.571)(1.5*10^-12,112.575)(1.4*10^-12,116.493)(1.3*10^-12,120.327)(1.2*10^-12,124.082)(1.1*10^-12,127.765)(1.*10^-12,131.388)};
       \addlegendentry{TK};
         \addplot[mark=none,very thick,dotted,teal] coordinates {(1.*10^-11,0.0563382)(9.9*10^-12,0.0576259)(9.8*10^-12,0.0589374)(9.7*10^-12,0.0602717)(9.6*10^-12,0.0616277)(9.5*10^-12,0.0630035)(9.4*10^-12,0.0643967)(9.3*10^-12,0.0658045)(9.2*10^-12,0.0672228)(9.1*10^-12,0.0686466)(9.*10^-12,0.0700696)(8.9*10^-12,0.0714839)(8.8*10^-12,0.0728793)(8.7*10^-12,0.0742434)(8.6*10^-12,0.0755608)(8.5*10^-12,0.0768123)(8.4*10^-12,0.0779742)(8.3*10^-12,0.0790178)(8.2*10^-12,0.079908)(8.1*10^-12,0.0806022)(8.*10^-12,0.0810493)(7.9*10^-12,0.0811884)(7.8*10^-12,0.0809478)(7.7*10^-12,0.080244)(7.6*10^-12,0.0789817)(7.5*10^-12,0.0770546)(7.4*10^-12,0.0743479)(7.3*10^-12,0.0720577)(7.2*10^-12,0.0727698)(7.1*10^-12,0.076026)(7.*10^-12,0.0836157)(6.9*10^-12,0.0931962)(6.8*10^-12,0.105345)(6.7*10^-12,0.120818)(6.6*10^-12,0.140595)(6.5*10^-12,0.165949)(6.4*10^-12,0.198519)(6.3*10^-12,0.240399)(6.2*10^-12,0.294236)(6.1*10^-12,0.363336)(6.*10^-12,0.451772)(5.9*10^-12,0.564483)(5.8*10^-12,0.707358)(5.7*10^-12,0.887291)(5.6*10^-12,1.11217)(5.5*10^-12,1.39084)(5.4*10^-12,1.73286)(5.3*10^-12,2.14828)(5.2*10^-12,2.64708)(5.1*10^-12,3.23859)(5.*10^-12,3.93062)(4.9*10^-12,4.72851)(4.8*10^-12,5.63423)(4.7*10^-12,6.6455)(4.6*10^-12,7.75533)(4.5*10^-12,8.95194)(4.4*10^-12,10.2194)(4.3*10^-12,11.5387)(4.2*10^-12,12.889)(4.1*10^-12,14.25)(4.*10^-12,15.6026)(3.9*10^-12,16.9309)(3.8*10^-12,18.2223)(3.7*10^-12,19.4685)(3.6*10^-12,20.6646)(3.5*10^-12,21.809)(3.4*10^-12,22.9028)(3.3*10^-12,23.949)(3.2*10^-12,24.952)(3.1*10^-12,25.9168)(3.*10^-12,26.849)(2.9*10^-12,27.7539)(2.8*10^-12,28.637)(2.7*10^-12,29.5031)(2.6*10^-12,30.3569)(2.5*10^-12,31.2025)(2.4*10^-12,32.0436)(2.3*10^-12,32.8835)(2.2*10^-12,33.7252)(2.1*10^-12,34.5711)(2.*10^-12,35.4235)(1.9*10^-12,36.2844)(1.8*10^-12,37.1555)(1.7*10^-12,38.0384)(1.6*10^-12,38.9345)(1.5*10^-12,39.8448)(1.4*10^-12,40.7706)(1.3*10^-12,41.7127)(1.2*10^-12,42.6721)(1.1*10^-12,43.6496)(1.*10^-12,44.646)        
        };
  \addlegendentry{TUR};
        \addplot[mark=none,thick,dashdotted,red] coordinates {
(1.*10^-11,3.36884)(9.9*10^-12,3.40715)(9.8*10^-12,3.4457)(9.7*10^-12,3.4844)(9.6*10^-12,3.52321)(9.5*10^-12,3.56202)(9.4*10^-12,3.60073)(9.3*10^-12,3.6392)(9.2*10^-12,3.67729)(9.1*10^-12,3.71481)(9.*10^-12,3.75152)(8.9*10^-12,3.78719)(8.8*10^-12,3.82149)(8.7*10^-12,3.85411)(8.6*10^-12,3.88465)(8.5*10^-12,3.9127)(8.4*10^-12,3.93786)(8.3*10^-12,3.95968)(8.2*10^-12,3.97778)(8.1*10^-12,3.99185)(8.*10^-12,4.00171)(7.9*10^-12,4.00738)(7.8*10^-12,4.00912)(7.7*10^-12,4.00753)(7.6*10^-12,4.00359)(7.5*10^-12,3.99868)(7.4*10^-12,3.99458)(7.3*10^-12,3.99347)(7.2*10^-12,3.99782)(7.1*10^-12,4.0103)(7.*10^-12,4.03373)(6.9*10^-12,4.07093)(6.8*10^-12,4.12465)(6.7*10^-12,4.19745)(6.6*10^-12,4.29167)(6.5*10^-12,4.40936)(6.4*10^-12,4.55216)(6.3*10^-12,4.72129)(6.2*10^-12,4.91749)(6.1*10^-12,5.14099)(6.*10^-12,5.39153)(5.9*10^-12,5.66849)(5.8*10^-12,5.971)(5.7*10^-12,6.29814)(5.6*10^-12,6.64923)(5.5*10^-12,7.02403)(5.4*10^-12,7.42299)(5.3*10^-12,7.84744)(5.2*10^-12,8.2996)(5.1*10^-12,8.78268)(5.*10^-12,9.30065)(4.9*10^-12,9.85807)(4.8*10^-12,10.4597)(4.7*10^-12,11.11)(4.6*10^-12,11.8128)(4.5*10^-12,12.5702)(4.4*10^-12,13.3827)(4.3*10^-12,14.2483)(4.2*10^-12,15.1624)(4.1*10^-12,16.1183)(4.*10^-12,17.1071)(3.9*10^-12,18.1187)(3.8*10^-12,19.1426)(3.7*10^-12,20.1689)(3.6*10^-12,21.1888)(3.5*10^-12,22.1953)(3.4*10^-12,23.1834)(3.3*10^-12,24.15)(3.2*10^-12,25.0942)(3.1*10^-12,26.0162)(3.*10^-12,26.9175)(2.9*10^-12,27.8006)(2.8*10^-12,28.6683)(2.7*10^-12,29.5239)(2.6*10^-12,30.3704)(2.5*10^-12,31.2111)(2.4*10^-12,32.0489)(2.3*10^-12,32.8867)(2.2*10^-12,33.7269)(2.1*10^-12,34.572)(2.*10^-12,35.4239)(1.9*10^-12,36.2846)(1.8*10^-12,37.1556)(1.7*10^-12,38.0385)(1.6*10^-12,38.9345)(1.5*10^-12,39.8449)(1.4*10^-12,40.7708)(1.3*10^-12,41.713)(1.2*10^-12,42.6725)(1.1*10^-12,43.6501)(1.*10^-12,44.6466)     
    };
  \addlegendentry{TUR (hyperaccurate)};
\end{axis}
\end{tikzpicture}
%%%
\begin{tikzpicture}
\begin{axis}[xlabel={$|F|$ [pN]},
    legend style={at={(0.05,0.75)},anchor=west}]
     \node[draw] at (0, 0)   (a) {(c)};
     \addplot[mark=none,thick,solid,black] coordinates {(1.*10^-11,134.653)(9.9*10^-12,135.641)(9.8*10^-12,136.633)(9.7*10^-12,137.626)(9.6*10^-12,138.619)(9.5*10^-12,139.613)(9.4*10^-12,140.605)(9.3*10^-12,141.595)(9.2*10^-12,142.58)(9.1*10^-12,143.559)(9.*10^-12,144.529)(8.9*10^-12,145.488)(8.8*10^-12,146.432)(8.7*10^-12,147.359)(8.6*10^-12,148.263)(8.5*10^-12,149.141)(8.4*10^-12,149.985)(8.3*10^-12,150.79)(8.2*10^-12,151.549)(8.1*10^-12,152.253)(8.*10^-12,152.893)(7.9*10^-12,153.46)(7.8*10^-12,153.941)(7.7*10^-12,154.327)(7.6*10^-12,154.606)(7.5*10^-12,154.765)(7.4*10^-12,154.797)(7.3*10^-12,154.691)(7.2*10^-12,154.445)(7.1*10^-12,154.059)(7.*10^-12,153.54)(6.9*10^-12,152.907)(6.8*10^-12,152.188)(6.7*10^-12,151.425)(6.6*10^-12,150.677)(6.5*10^-12,150.015)(6.4*10^-12,149.527)(6.3*10^-12,149.312)(6.2*10^-12,149.472)(6.1*10^-12,150.113)(6.*10^-12,151.332)(5.9*10^-12,153.212)(5.8*10^-12,155.818)(5.7*10^-12,159.19)(5.6*10^-12,163.346)(5.5*10^-12,168.28)(5.4*10^-12,173.968)(5.3*10^-12,180.371)(5.2*10^-12,187.441)(5.1*10^-12,195.126)(5.*10^-12,203.372)(4.9*10^-12,212.13)(4.8*10^-12,221.354)(4.7*10^-12,231.004)(4.6*10^-12,241.046)(4.5*10^-12,251.452)(4.4*10^-12,262.202)(4.3*10^-12,273.277)(4.2*10^-12,284.667)(4.1*10^-12,296.363)(4.*10^-12,308.359)(3.9*10^-12,320.655)(3.8*10^-12,333.248)(3.7*10^-12,346.141)(3.6*10^-12,359.336)(3.5*10^-12,372.837)(3.4*10^-12,386.648)(3.3*10^-12,400.774)(3.2*10^-12,415.22)(3.1*10^-12,429.993)(3.*10^-12,445.097)(2.9*10^-12,460.54)(2.8*10^-12,476.326)(2.7*10^-12,492.462)(2.6*10^-12,508.955)(2.5*10^-12,525.81)(2.4*10^-12,543.033)(2.3*10^-12,560.631)(2.2*10^-12,578.61)(2.1*10^-12,596.975)(2.*10^-12,615.734)(1.9*10^-12,634.89)(1.8*10^-12,654.452)(1.7*10^-12,674.424)(1.6*10^-12,694.812)(1.5*10^-12,715.623)(1.4*10^-12,736.862)(1.3*10^-12,758.535)(1.2*10^-12,780.648)(1.1*10^-12,803.208)(1.*10^-12,826.218)
     };
      %      \addlegendentry{$\Delta \sigma$};
	 \addplot[mark=none,thick,dashed,blue] coordinates {
      (1.*10^-11,4.24223)(9.9*10^-12,4.2751)(9.8*10^-12,4.30809)(9.7*10^-12,4.34119)(9.6*10^-12,4.37437)(9.5*10^-12,4.40757)(9.4*10^-12,4.44075)(9.3*10^-12,4.47385)(9.2*10^-12,4.50678)(9.1*10^-12,4.53946)(9.*10^-12,4.57177)(8.9*10^-12,4.60356)(8.8*10^-12,4.63467)(8.7*10^-12,4.66488)(8.6*10^-12,4.69394)(8.5*10^-12,4.72155)(8.4*10^-12,4.76444)(8.3*10^-12,4.83075)(8.2*10^-12,4.9047)(8.1*10^-12,4.98807)(8.*10^-12,5.08304)(7.9*10^-12,5.19231)(7.8*10^-12,5.31926)(7.7*10^-12,5.46806)(7.6*10^-12,5.64392)(7.5*10^-12,5.8533)(7.4*10^-12,6.10426)(7.3*10^-12,6.4068)(7.2*10^-12,6.77339)(7.1*10^-12,7.2195)(7.*10^-12,7.7643)(6.9*10^-12,8.43142)(6.8*10^-12,9.24981)(6.7*10^-12,10.2546)(6.6*10^-12,11.4879)(6.5*10^-12,12.9992)(6.4*10^-12,14.8459)(6.3*10^-12,17.0924)(6.2*10^-12,19.8089)(6.1*10^-12,23.0694)(6.*10^-12,26.948)(5.9*10^-12,31.5154)(5.8*10^-12,36.8348)(5.7*10^-12,42.9577)(5.6*10^-12,49.9212)(5.5*10^-12,57.7464)(5.4*10^-12,66.438)(5.3*10^-12,75.9861)(5.2*10^-12,86.3679)(5.1*10^-12,97.5511)(5.*10^-12,109.497)(4.9*10^-12,122.163)(4.8*10^-12,135.506)(4.7*10^-12,149.481)(4.6*10^-12,164.048)(4.5*10^-12,179.165)(4.4*10^-12,194.792)(4.3*10^-12,210.89)(4.2*10^-12,227.419)(4.1*10^-12,244.338)(4.*10^-12,261.601)(3.9*10^-12,279.158)(3.8*10^-12,296.954)(3.7*10^-12,314.926)(3.6*10^-12,333.005)(3.5*10^-12,351.113)(3.4*10^-12,369.167)(3.3*10^-12,387.077)(3.2*10^-12,404.753)(3.1*10^-12,422.106)(3.*10^-12,439.054)(2.9*10^-12,455.523)(2.8*10^-12,471.456)(2.7*10^-12,486.811)(2.6*10^-12,501.569)(2.5*10^-12,515.726)(2.4*10^-12,529.299)(2.3*10^-12,542.318)(2.2*10^-12,554.825)(2.1*10^-12,566.87)(2.*10^-12,578.507)(1.9*10^-12,589.79)(1.8*10^-12,600.774)(1.7*10^-12,611.509)(1.6*10^-12,622.042)(1.5*10^-12,632.415)(1.4*10^-12,642.667)(1.3*10^-12,652.828)(1.2*10^-12,662.929)(1.1*10^-12,672.993)(1.*10^-12,683.041)
         };
  %     \addlegendentry{TK};
         \addplot[mark=none,very thick,dotted,teal] coordinates {(1.*10^-11,4.21274)(9.9*10^-12,4.24537)(9.8*10^-12,4.27814)(9.7*10^-12,4.31101)(9.6*10^-12,4.34396)(9.5*10^-12,4.37693)(9.4*10^-12,4.40989)(9.3*10^-12,4.44276)(9.2*10^-12,4.47548)(9.1*10^-12,4.50794)(9.*10^-12,4.54003)(8.9*10^-12,4.57161)(8.8*10^-12,4.60252)(8.7*10^-12,4.63254)(8.6*10^-12,4.66143)(8.5*10^-12,4.68887)(8.4*10^-12,4.73138)(8.3*10^-12,4.79698)(8.2*10^-12,4.8701)(8.1*10^-12,4.95248)(8.*10^-12,5.04626)(7.9*10^-12,5.15412)(7.8*10^-12,5.27933)(7.7*10^-12,5.426)(7.6*10^-12,5.59922)(7.5*10^-12,5.80528)(7.4*10^-12,6.05203)(7.3*10^-12,6.34919)(7.2*10^-12,6.70881)(7.1*10^-12,7.14578)(7.*10^-12,7.67844)(6.9*10^-12,8.32922)(6.8*10^-12,9.12533)(6.7*10^-12,10.0993)(6.6*10^-12,11.2895)(6.5*10^-12,12.7399)(6.4*10^-12,14.4996)(6.3*10^-12,16.6209)(6.2*10^-12,19.1569)(6.1*10^-12,22.1569)(6.*10^-12,25.6613)(5.9*10^-12,29.6956)(5.8*10^-12,34.2645)(5.7*10^-12,39.3474)(5.6*10^-12,44.8964)(5.5*10^-12,50.8377)(5.4*10^-12,57.076)(5.3*10^-12,63.5028)(5.2*10^-12,70.0053)(5.1*10^-12,76.4759)(5.*10^-12,82.8201)(4.9*10^-12,88.9618)(4.8*10^-12,94.8457)(4.7*10^-12,100.437)(4.6*10^-12,105.721)(4.5*10^-12,110.697)(4.4*10^-12,115.376)(4.3*10^-12,119.778)(4.2*10^-12,123.928)(4.1*10^-12,127.855)(4.*10^-12,131.585)(3.9*10^-12,135.147)(3.8*10^-12,138.566)(3.7*10^-12,141.866)(3.6*10^-12,145.068)(3.5*10^-12,148.193)(3.4*10^-12,151.255)(3.3*10^-12,154.27)(3.2*10^-12,157.251)(3.1*10^-12,160.208)(3.*10^-12,163.15)(2.9*10^-12,166.085)(2.8*10^-12,169.021)(2.7*10^-12,171.962)(2.6*10^-12,174.913)(2.5*10^-12,177.88)(2.4*10^-12,180.864)(2.3*10^-12,183.87)(2.2*10^-12,186.898)(2.1*10^-12,189.953)(2.*10^-12,193.035)(1.9*10^-12,196.146)(1.8*10^-12,199.287)(1.7*10^-12,202.46)(1.6*10^-12,205.665)(1.5*10^-12,208.903)(1.4*10^-12,212.175)(1.3*10^-12,215.482)(1.2*10^-12,218.823)(1.1*10^-12,222.201)(1.*10^-12,225.614)        };
 % \addlegendentry{TUR};
  \addplot[mark=none,thick,dashdotted,red] coordinates {
        (1.*10^-11,17.1806)(9.9*10^-12,17.3188)(9.8*10^-12,17.4565)(9.7*10^-12,17.5936)(9.6*10^-12,17.7296)(9.5*10^-12,17.8644)(9.4*10^-12,17.9976)(9.3*10^-12,18.1289)(9.2*10^-12,18.2583)(9.1*10^-12,18.3857)(9.*10^-12,18.5113)(8.9*10^-12,18.6355)(8.8*10^-12,18.7589)(8.7*10^-12,18.8828)(8.6*10^-12,19.0085)(8.5*10^-12,19.138)(8.4*10^-12,19.2736)(8.3*10^-12,19.4182)(8.2*10^-12,19.5748)(8.1*10^-12,19.7471)(8.*10^-12,19.9391)(7.9*10^-12,20.1554)(7.8*10^-12,20.4011)(7.7*10^-12,20.6822)(7.6*10^-12,21.0055)(7.5*10^-12,21.379)(7.4*10^-12,21.812)(7.3*10^-12,22.3155)(7.2*10^-12,22.902)(7.1*10^-12,23.5859)(7.*10^-12,24.3836)(6.9*10^-12,25.3131)(6.8*10^-12,26.3941)(6.7*10^-12,27.6474)(6.6*10^-12,29.0947)(6.5*10^-12,30.7575)(6.4*10^-12,32.6568)(6.3*10^-12,34.812)(6.2*10^-12,37.2403)(6.1*10^-12,39.9559)(6.*10^-12,42.9694)(5.9*10^-12,46.2872)(5.8*10^-12,49.9103)(5.7*10^-12,53.8335)(5.6*10^-12,58.0436)(5.5*10^-12,62.5189)(5.4*10^-12,67.2275)(5.3*10^-12,72.1278)(5.2*10^-12,77.1693)(5.1*10^-12,82.2953)(5.*10^-12,87.4465)(4.9*10^-12,92.5648)(4.8*10^-12,97.5978)(4.7*10^-12,102.502)(4.6*10^-12,107.244)(4.5*10^-12,111.803)(4.4*10^-12,116.167)(4.3*10^-12,120.337)(4.2*10^-12,124.318)(4.1*10^-12,128.123)(4.*10^-12,131.767)(3.9*10^-12,135.268)(3.8*10^-12,138.646)(3.7*10^-12,141.918)(3.6*10^-12,145.101)(3.5*10^-12,148.213)(3.4*10^-12,151.267)(3.3*10^-12,154.277)(3.2*10^-12,157.254)(3.1*10^-12,160.209)(3.*10^-12,163.15)(2.9*10^-12,166.085)(2.8*10^-12,169.021)(2.7*10^-12,171.962)(2.6*10^-12,174.914)(2.5*10^-12,177.881)(2.4*10^-12,180.865)(2.3*10^-12,183.871)(2.2*10^-12,186.9)(2.1*10^-12,189.955)(2.*10^-12,193.037)(1.9*10^-12,196.149)(1.8*10^-12,199.29)(1.7*10^-12,202.463)(1.6*10^-12,205.668)(1.5*10^-12,208.907)(1.4*10^-12,212.179)(1.3*10^-12,215.486)(1.2*10^-12,218.828)(1.1*10^-12,222.205)(1.*10^-12,225.619)};
%  \addlegendentry{TUR (hyperaccurate)};
\end{axis}
\end{tikzpicture}
%%%
\begin{tikzpicture}
\begin{axis}[xlabel={$|F|$ [pN]},
    legend style={at={(0.05,0.75)},anchor=west}]
     \node[draw] at (0, 0)   (a) {(d)};
     \addplot[mark=none,thick,solid,black] coordinates {(1.*10^-11,611.241)(9.9*10^-12,615.914)(9.8*10^-12,620.604)(9.7*10^-12,625.306)(9.6*10^-12,630.016)(9.5*10^-12,634.728)(9.4*10^-12,639.435)(9.3*10^-12,644.13)(9.2*10^-12,648.801)(9.1*10^-12,653.437)(9.*10^-12,658.023)(8.9*10^-12,662.542)(8.8*10^-12,666.973)(8.7*10^-12,671.29)(8.6*10^-12,675.464)(8.5*10^-12,679.46)(8.4*10^-12,683.236)(8.3*10^-12,686.744)(8.2*10^-12,689.927)(8.1*10^-12,692.72)(8.*10^-12,695.046)(7.9*10^-12,696.822)(7.8*10^-12,697.953)(7.7*10^-12,698.336)(7.6*10^-12,697.859)(7.5*10^-12,696.411)(7.4*10^-12,693.88)(7.3*10^-12,690.164)(7.2*10^-12,685.183)(7.1*10^-12,678.886)(7.*10^-12,671.273)(6.9*10^-12,662.4)(6.8*10^-12,652.401)(6.7*10^-12,641.49)(6.6*10^-12,629.966)(6.5*10^-12,618.202)(6.4*10^-12,606.629)(6.3*10^-12,595.705)(6.2*10^-12,585.883)(6.1*10^-12,577.577)(6.*10^-12,571.13)(5.9*10^-12,566.797)(5.8*10^-12,564.732)(5.7*10^-12,564.997)(5.6*10^-12,567.564)(5.5*10^-12,572.343)(5.4*10^-12,579.193)(5.3*10^-12,587.944)(5.2*10^-12,598.415)(5.1*10^-12,610.42)(5.*10^-12,623.786)(4.9*10^-12,638.35)(4.8*10^-12,653.969)(4.7*10^-12,670.515)(4.6*10^-12,687.881)(4.5*10^-12,705.975)(4.4*10^-12,724.719)(4.3*10^-12,744.052)(4.2*10^-12,763.921)(4.1*10^-12,784.284)(4.*10^-12,805.109)(3.9*10^-12,826.368)(3.8*10^-12,848.042)(3.7*10^-12,870.113)(3.6*10^-12,892.571)(3.5*10^-12,915.405)(3.4*10^-12,938.609)(3.3*10^-12,962.179)(3.2*10^-12,986.113)(3.1*10^-12,1010.41)(3.*10^-12,1035.06)(2.9*10^-12,1060.08)(2.8*10^-12,1085.46)(2.7*10^-12,1111.21)(2.6*10^-12,1137.32)(2.5*10^-12,1163.8)(2.4*10^-12,1190.65)(2.3*10^-12,1217.88)(2.2*10^-12,1245.49)(2.1*10^-12,1273.47)(2.*10^-12,1301.85)(1.9*10^-12,1330.61)(1.8*10^-12,1359.76)(1.7*10^-12,1389.31)(1.6*10^-12,1419.26)(1.5*10^-12,1449.62)(1.4*10^-12,1480.38)(1.3*10^-12,1511.56)(1.2*10^-12,1543.15)(1.1*10^-12,1575.16)(1.*10^-12,1607.6)
     };
  %     \addlegendentry{$\Delta \sigma$};
         \addplot[mark=none,thick,dashed,blue] coordinates { 
         (1.*10^-11,93.7751)(9.9*10^-12,94.4849)(9.8*10^-12,95.1935)(9.7*10^-12,95.8995)(9.6*10^-12,96.6009)(9.5*10^-12,97.2953)(9.4*10^-12,97.9798)(9.3*10^-12,98.6508)(9.2*10^-12,99.3196)(9.1*10^-12,100.162)(9.*10^-12,101.027)(8.9*10^-12,101.918)(8.8*10^-12,102.839)(8.7*10^-12,103.797)(8.6*10^-12,104.798)(8.5*10^-12,105.852)(8.4*10^-12,106.969)(8.3*10^-12,108.162)(8.2*10^-12,109.446)(8.1*10^-12,110.842)(8.*10^-12,112.373)(7.9*10^-12,114.068)(7.8*10^-12,115.96)(7.7*10^-12,118.093)(7.6*10^-12,120.513)(7.5*10^-12,123.282)(7.4*10^-12,126.466)(7.3*10^-12,130.145)(7.2*10^-12,134.409)(7.1*10^-12,139.362)(7.*10^-12,145.115)(6.9*10^-12,151.788)(6.8*10^-12,159.507)(6.7*10^-12,168.397)(6.6*10^-12,178.577)(6.5*10^-12,190.154)(6.4*10^-12,203.213)(6.3*10^-12,217.813)(6.2*10^-12,233.982)(6.1*10^-12,251.712)(6.*10^-12,270.963)(5.9*10^-12,291.666)(5.8*10^-12,313.723)(5.7*10^-12,337.022)(5.6*10^-12,361.435)(5.5*10^-12,386.829)(5.4*10^-12,413.069)(5.3*10^-12,440.017)(5.2*10^-12,467.539)(5.1*10^-12,495.498)(5.*10^-12,523.76)(4.9*10^-12,552.184)(4.8*10^-12,580.628)(4.7*10^-12,608.944)(4.6*10^-12,636.98)(4.5*10^-12,664.577)(4.4*10^-12,691.577)(4.3*10^-12,717.827)(4.2*10^-12,743.179)(4.1*10^-12,767.504)(4.*10^-12,790.695)(3.9*10^-12,812.674)(3.8*10^-12,833.396)(3.7*10^-12,852.848)(3.6*10^-12,871.052)(3.5*10^-12,888.059)(3.4*10^-12,903.941)(3.3*10^-12,918.791)(3.2*10^-12,932.708)(3.1*10^-12,945.798)(3.*10^-12,958.166)(2.9*10^-12,969.911)(2.8*10^-12,981.125)(2.7*10^-12,991.893)(2.6*10^-12,1002.29)(2.5*10^-12,1012.38)(2.4*10^-12,1022.22)(2.3*10^-12,1031.87)(2.2*10^-12,1041.35)(2.1*10^-12,1050.71)(2.*10^-12,1059.98)(1.9*10^-12,1069.18)(1.8*10^-12,1078.33)(1.7*10^-12,1087.45)(1.6*10^-12,1096.55)(1.5*10^-12,1105.65)(1.4*10^-12,1114.76)(1.3*10^-12,1123.88)(1.2*10^-12,1133.01)(1.1*10^-12,1142.18)(1.*10^-12,1151.38)
         };
   %    \addlegendentry{TK};
         \addplot[mark=none,very thick,dotted,teal] coordinates {
(1.*10^-11,79.8409)(9.9*10^-12,80.4489)(9.8*10^-12,81.0569)(9.7*10^-12,81.6637)(9.6*10^-12,82.268)(9.5*10^-12,82.8679)(9.4*10^-12,83.4615)(9.3*10^-12,84.046)(9.2*10^-12,84.6295)(9.1*10^-12,85.3368)(9.*10^-12,86.0602)(8.9*10^-12,86.8024)(8.8*10^-12,87.5664)(8.7*10^-12,88.3563)(8.6*10^-12,89.1769)(8.5*10^-12,90.0339)(8.4*10^-12,90.9345)(8.3*10^-12,91.8876)(8.2*10^-12,92.9036)(8.1*10^-12,93.9954)(8.*10^-12,95.1785)(7.9*10^-12,96.4714)(7.8*10^-12,97.8961)(7.7*10^-12,99.4786)(7.6*10^-12,101.249)(7.5*10^-12,103.243)(7.4*10^-12,105.499)(7.3*10^-12,108.061)(7.2*10^-12,110.976)(7.1*10^-12,114.292)(7.*10^-12,118.059)(6.9*10^-12,122.318)(6.8*10^-12,127.106)(6.7*10^-12,132.443)(6.6*10^-12,138.33)(6.5*10^-12,144.746)(6.4*10^-12,151.643)(6.3*10^-12,158.945)(6.2*10^-12,166.554)(6.1*10^-12,174.356)(6.*10^-12,182.225)(5.9*10^-12,190.041)(5.8*10^-12,197.694)(5.7*10^-12,205.091)(5.6*10^-12,212.163)(5.5*10^-12,218.865)(5.4*10^-12,225.174)(5.3*10^-12,231.087)(5.2*10^-12,236.615)(5.1*10^-12,241.781)(5.*10^-12,246.615)(4.9*10^-12,251.151)(4.8*10^-12,255.423)(4.7*10^-12,259.467)(4.6*10^-12,263.314)(4.5*10^-12,266.995)(4.4*10^-12,270.535)(4.3*10^-12,273.96)(4.2*10^-12,277.289)(4.1*10^-12,280.542)(4.*10^-12,283.732)(3.9*10^-12,286.874)(3.8*10^-12,289.979)(3.7*10^-12,293.057)(3.6*10^-12,296.115)(3.5*10^-12,299.16)(3.4*10^-12,302.198)(3.3*10^-12,305.234)(3.2*10^-12,308.272)(3.1*10^-12,311.315)(3.*10^-12,314.366)(2.9*10^-12,317.427)(2.8*10^-12,320.501)(2.7*10^-12,323.589)(2.6*10^-12,326.693)(2.5*10^-12,329.814)(2.4*10^-12,332.953)(2.3*10^-12,336.11)(2.2*10^-12,339.288)(2.1*10^-12,342.485)(2.*10^-12,345.704)(1.9*10^-12,348.944)(1.8*10^-12,352.206)(1.7*10^-12,355.49)(1.6*10^-12,358.798)(1.5*10^-12,362.128)(1.4*10^-12,365.482)(1.3*10^-12,368.859)(1.2*10^-12,372.261)(1.1*10^-12,375.687)(1.*10^-12,379.138)    };
  %\addlegendentry{TUR};
         \addplot[mark=none,thick,dashdotted,red] coordinates {
         (1.*10^-11,93.1931)(9.9*10^-12,93.9067)(9.8*10^-12,94.6232)(9.7*10^-12,95.3429)(9.6*10^-12,96.0663)(9.5*10^-12,96.7942)(9.4*10^-12,97.5278)(9.3*10^-12,98.2686)(9.2*10^-12,99.0184)(9.1*10^-12,99.7796)(9.*10^-12,100.555)(8.9*10^-12,101.347)(8.8*10^-12,102.159)(8.7*10^-12,102.995)(8.6*10^-12,103.859)(8.5*10^-12,104.756)(8.4*10^-12,105.692)(8.3*10^-12,106.673)(8.2*10^-12,107.709)(8.1*10^-12,108.808)(8.*10^-12,109.982)(7.9*10^-12,111.246)(7.8*10^-12,112.616)(7.7*10^-12,114.113)(7.6*10^-12,115.76)(7.5*10^-12,117.584)(7.4*10^-12,119.617)(7.3*10^-12,121.894)(7.2*10^-12,124.452)(7.1*10^-12,127.332)(7.*10^-12,130.574)(6.9*10^-12,134.217)(6.8*10^-12,138.294)(6.7*10^-12,142.83)(6.6*10^-12,147.835)(6.5*10^-12,153.306)(6.4*10^-12,159.216)(6.3*10^-12,165.518)(6.2*10^-12,172.145)(6.1*10^-12,179.012)(6.*10^-12,186.021)(5.9*10^-12,193.07)(5.8*10^-12,200.06)(5.7*10^-12,206.902)(5.6*10^-12,213.522)(5.5*10^-12,219.866)(5.4*10^-12,225.898)(5.3*10^-12,231.601)(5.2*10^-12,236.974)(5.1*10^-12,242.028)(5.*10^-12,246.782)(4.9*10^-12,251.261)(4.8*10^-12,255.495)(4.7*10^-12,259.512)(4.6*10^-12,263.342)(4.5*10^-12,267.011)(4.4*10^-12,270.545)(4.3*10^-12,273.965)(4.2*10^-12,277.292)(4.1*10^-12,280.542)(4.*10^-12,283.732)(3.9*10^-12,286.875)(3.8*10^-12,289.98)(3.7*10^-12,293.058)(3.6*10^-12,296.116)(3.5*10^-12,299.162)(3.4*10^-12,302.2)(3.3*10^-12,305.237)(3.2*10^-12,308.275)(3.1*10^-12,311.318)(3.*10^-12,314.37)(2.9*10^-12,317.431)(2.8*10^-12,320.506)(2.7*10^-12,323.594)(2.6*10^-12,326.698)(2.5*10^-12,329.819)(2.4*10^-12,332.958)(2.3*10^-12,336.116)(2.2*10^-12,339.294)(2.1*10^-12,342.492)(2.*10^-12,345.71)(1.9*10^-12,348.951)(1.8*10^-12,352.213)(1.7*10^-12,355.497)(1.6*10^-12,358.805)(1.5*10^-12,362.135)(1.4*10^-12,365.489)(1.3*10^-12,368.867)(1.2*10^-12,372.269)(1.1*10^-12,375.695)(1.*10^-12,379.146) };
%  \addlegendentry{TUR (hyperaccurate)};
\end{axis}
\end{tikzpicture}
\caption{Comparison of the  for the entropy production (solid) for the kinesin model in the stationary state. For $f(e)=\text{sign}(p(e)-p(\bar e))$, %$f(e)=|p(e)-p(\bar e)|$, 
the new bound Eq. \eqref{eq:crsigmahantisym} (dashed) and the thermodynamic uncertainty relation Eq.~\eqref{eq:classicTUR} (dotted). For the hyper-accurate current $f^*(e)=(p(e)-p(\bar e))/(p(e)+p(\bar e))$ (making Eq.~\eqref{eq:classicTUR} as tight as can be, see Appendix), the thermodynamic uncertainty relation Eq.~\eqref{eq:classicTUR} (dashed-dotted). From panel (a) to (d) the ATP concentration increased as $\text{[ATP]}=1,10,10^2,10^3$.}
\label{fig:kinesin}
\end{figure*}

\section{Discussion and applications}

\label{sec:5}

\subsection{Tightness of Thermo-Kinetic relations and the interest of non-conservative forces}  It is known \cite{dechant2022minimum} that the classical Speed Limit \eqref{eq:cslactfull} is tight for full entropy production and activity observable. 
%We provide an elementary proof in the Appendix. 
This means that given $p_\text{ini}$, $p_\text{fin}$ and a given `budget' for the expected number of jumps one can find a time-varying Markov process that drives  $p_\text{ini}$ to $p_\text{fin}$ within the prescribed number of jumps whose entropy production is the r.h.s. of \eqref{eq:cslactfull}. Remarkably, this optimal strategy only involves conservative forces, i.e. has zero housekeeping entropy production ($\Delta \sigma= \Delta \sigma_{NA}$). 

Comparing \eqref{eq:sigmaHS-h-TV} to \eqref{eq:cslactfull}, we observe that the l.h.s.  of the former   is lower or equal to the l.h.s. of the latter ($\Delta \sigma_{NA} \leq \Delta \sigma$), and the same for the r.h.s. ($h \leq h^*_\text{sym}$). We can thus wonder how tight \eqref{eq:sigmaHS-h-TV} is, as a bound on $\Delta \sigma_{NA}$. In particular, can we find a Markov chain with $\Delta \sigma_{NA}$  strictly smaller than the r.h.s of \eqref{eq:sigma-hsym}? It turns out that we can.  For example, a simple three-state Markov chain running for a short time $dt$ with clockwise outflows $25dt$, $5 dt$, $1 dt$ (and no counterclockwise flows), has $\langle A \rangle=31 dt$,  $d_{TV}(p_{X_\text{ini}},p_{X_\text{fin}})/\langle A \rangle=24/31=0.77$ and $\Delta \sigma_{NA}=45.1 dt < \langle A \rangle h^*_\text{sym}(0.77)=49.5 dt$. This means that, albeit minimum entropy production for a given trajectory is reached with a conservative force protocol, it is possible to reach an even lower NA entropy production with a non-conservative protocol (thus at the cost of higher total entropy production). Thus there is a trade-off between NA and housekeeping entropy production, as in some cases the latter can be increased in order to decrease the former. Said otherwise, non-equilibrium stationary states can be used to achieve faster transitions, for a given budget in $\Delta \sigma_{NA}$.

Nevertheless the range of this trade-off is quite limited. Indeed $h^*_\text{sym}$ and $h^*$ always differ in value by less than 18\%, in either direction. This gap goes to 0\% both for small arguments (ratio $d_{TV}/\int \langle A \rangle dt \ll 1$, slow kinetics),  and large arguments (ratio $d_{TV}/\int \langle A \rangle dt$ close to one, fast kinetics). Thus the trade-off exists in an intermediate range of speeds. 

\subsection{Constant or time symmetric rates revisited}
Let us now focus again on constant-rate (or time-symmetric-rate) Markov chains, in the case of activity observable. Weakening  \eqref{eq:constantfullacti} with $\langle A \rangle \leq 1$ (as $\langle A \rangle$ is a probability), we find 
\begin{align}
	\Delta \sigma  + D(p_{X_\text{fin}}\|p_{X_\text{ini}})& \geq \langle A \rangle  h^*_\text{sym}\left(\frac{d_{TV}(p_{X_\text{ini}},p_{X_\text{fin}})}{\langle A \rangle}\right) \\
	&\geq h^*_\text{sym}(d_{TV}(p_{X_\text{ini}},p_{X_\text{fin}})).%+h(d_{TV}(p_{X_\text{ini}},p_{X_\text{fin}}))
	%\\
	%	&\geq (1+d_{TV}(p_{X_\text{ini}},p_{X_\text{fin}})) \ln \frac{1}{1-d_{TV}(p_{X_\text{ini}},p_{X_\text{fin}})}\\ &\quad - \ln (1+d_{TV}(p_{X_\text{ini}},p_{X_\text{fin}}))
\end{align}
%where $h$ is taken here as Gilardoni's function.
Interestingly, this implies that for those initial and final state distributions such that $h^*(d_{TV}(p_{X_\text{ini}},p_{X_\text{fin}})) \leq D(p_{X_\text{fin}}\|p_{X_\text{ini}}) < h^*_\text{sym}(d_{TV}(p_{X_\text{ini}},p_{X_\text{fin}}))$, there is an incompressible  lower bound for entropy production regardless of the activity or duration of the transformation. 
%\textcolor{red}{NUMERICAL EXAMPLE?? Eg Passing from $(0.8,0.2)$ to  $(0.3,0.7)$ with constant rates (RANDOM numbers, find correct ones) must cost at least $\Delta \sigma\geq ????$}

Finally a word on the Markovian assumption here. The Thermo-Kinetic relations \eqref{eq:crsigmahsymD},\eqref{eq:crsigmahantisym} and their variants are all lower bounds on $D(p\|\tilde{p})$. The condition of constant or time-symmetric Markov chains is sufficient to ensure that $D(p\|\tilde{p})= \Delta \sigma + D(p_{X_{\text{ini}}},p_{X_{\text{fin}}})$, which provides a physical meaning to the bound. This relation however is also true in some non-Markovian contexts, e.g. in presence of feedback \cite{potts2019thermodynamic,van2020uncertainty}, similarly for the relations \eqref{eq:constsigmaNAantisym} and \eqref{eq:constsigmaNAtransport}.

\subsection{The molecular motor kinesin} We illustrate the bounds on a toy model describing the motion of the kinesin molecular motor along a microtubule powered by ATP hydrolysis and exerting a force $F$ on a pulled cargo. This standard model is a two-state, eight-transition Markov chain at a non-equilibrium stationarity. We refer to \cite{dit18,falasco2020unifying} for a detailed description of the model. In Fig.~\eqref{fig:kinesin} we compare the Thermo-Kinetic bound Eq. \eqref{eq:sigma-hsym} and the standard Thermodynamic Uncertainty Relation for various values of ATP concentration and pulling force $F$, and compare the quality of the bounds on entropy production. We see in particular that \eqref{eq:sigma-hsym} tends to compare favorably to the Thermodynamic Uncertainty Relation far from equilibrium (high ATP concentration).

\subsection{Switching a bit in an electronic memory}
 A logical bit (zero or one) can be encoded in simple or complex nonlinear circuit, depending on the functionality (NOT gate, memory, clock, etc.). In  generic circumstances, such circuits can be modelled by an overdamped Markov chain, the states of which record the charge (number of electrons) present in each conductor of the circuit (typically the plate of a capacitance) \cite{freitas2020stochastic,freitas2021linear}. The circuit state is in probability distribution $p_0$ when encoding a logical zero, and in probability distribution $p_1$ when encoding a one. The switching from a zero to a one thus amounts to driving the system from $p_\text{ini}=p_0$ to $p_\text{fin}=p_1$. 
The value of the bit is retrieved by the reading of the charge (or equivalently, voltage) at the plate of a single `output' capacitance $C_\text{out}$. The distribution $p_0$ (respectively, $p_1$) results in a mean charge $q_0$ (respectively, $q_1$) on $C_\text{out}$. 

We now use the Thermo-Kinetic Relation to obtain a lower bound on the entropy produced while switching the system from a zero to a one, i.e. from $p_\text{ini}=p_0$ to $p_\text{fin}=p_1$. Note that Landauer's cost here is zero, as we do not erase or reset a bit but simply flip it.

The antisymmetric observable $f=\Delta q$ of interest on transitions is simply the `gradient' of charge, i.e. the increase of charge into the positive plate of $C_\text{out}$ along the transition. Thus $f=\pm q_e$ or $0$, where $q_e$ is the (positive) charge of the electron. It is possible that many transitions are zero valued, because they do not represent a change of charge on $C_\text{out}$ (but on other parts of the complete circuit). In this way, the absolute value along the trajectory $\int \langle |f| \rangle dt$, counting  the number of charge carriers flowing in or out the capacitance $C_\text{out}$, is less than the total activity of the full circuit along the trajectory. The net change of charge over the trajectory $\int \langle f \rangle dt$ is simply $q_1-q_0$, the total charge separating a zero from a one. The maximum value for $f$ over an infinitesimal transition is $q_e$. Thus we find 
\begin{align}
	\Delta \sigma \geq 2 \frac{|q_1-q_0|}{q_e} \text{atanh}\frac{|q_1-q_0|}{\int |\Delta q| dt}.
\end{align}
The quantity $\int |\Delta q| dt/ q_e$ can be considered as a measure of the time taken by the driving. Indeed, a slower driving offers more opportunities for positive or negative charge variation in the output capacitance.  In particular we recover Pinsker's bound, with a bound inversely proportional to time, consistently e.g. with \cite{aurell2011optimal,lee2022speed} in other contexts:
\begin{align}\label{eq:pinskerbitflip}
	\Delta \sigma \geq 2 \frac{|q_1-q_0|^2}{q_e \int |\Delta q| dt}.
\end{align}
At stationarity, the number of charge into the positive plate of $C_\text{out}$ match (on average)  the number of charge out of that same plate. Thus $f$ verifies the conditions to apply the non-adiabatic Thermo-Kinetic Relation:
\begin{align}
	\Delta \sigma_{NA} \geq 2 \frac{\int |\Delta q| dt}{q_e} h\left(\frac{|q_1-q_0|}{\int |\Delta q| dt}\right).
\end{align}
This relaxes in particular to the Pinsker's bound, stronger than \eqref{eq:pinskerbitflip} since it applies to the sole non-adiabatic  entropy production:
\begin{align}\label{eq:pinskerNAbitflip}
	\Delta \sigma_{NA} \geq 2 \frac{|q_1-q_0|^2}{q_e \int |\Delta q| dt}.
\end{align}
This is a practical bound that can assess the efficiency of a wide range of electronic devices in storing and writing a bit. It can easily be adapted to non-electronic overdamped devices as well (e.g. in chemical computing).

\section{Acknowledgments}
J-C. D. acknowledges funding from the Project INTER/FNRS/20/15074473 `TheCirco' on Thermodynamics of Circuits for Computation, funded by the F.R.S.-FNRS (Belgium) and FNR (Luxembourg).
G.F. is funded by the European Union -- NextGenerationEU -- and by the program STARS@UNIPD with project ``ThermoComplex".

\bibliography{references}

\appendix

\section{Effective maximum in bounds between Kullback-Leilber Divergences}
 We seek to improve \eqref{eq:DpqA} and \eqref{eq:Dpqphi} by replacing $\varphi_{\max}$ and $A_{\max}$ with an `effective' maximum, whenever  $\varphi$ (thus $A$) is unbounded.
 We can safely assume that $D(p\|q)$ is finite, otherwise \eqref{eq:DpqA} and \eqref{eq:Dpqphi} are trivially true. Thus the sum \eqref{eq:sumDphi}, although gathering potentially infinitely many positive  terms, converges, thus the sum of all terms for which $z > \varphi_0$, for a large enough treshold $\varphi_0$, is as small as we desire. Thus for any $0 < \epsilon < 1$ there exists a $\varphi_0$ such that 
\begin{align}p(\varphi > \varphi_0)& D( \varphi p\| \varphi  q | \varphi > \varphi_M) \\
	&= \sum_{z >\varphi_0} zp(\varphi=z)  D(\varphi p\|\varphi  q | \varphi=z) \\ &< \epsilon D(\varphi p\|\varphi  q).
\end{align}
Then we find that  
\begin{align}
	D(\varphi p\|\varphi  q) &=  p(\varphi \leq \varphi_0) D( \varphi p\| \varphi  q | \varphi \leq \varphi_0) \nonumber\\&\;\;\;+ p(\varphi > \varphi_0) D( \varphi p\| \varphi  q | \varphi > \varphi_0) \\
	&\leq \varphi_0 p(\varphi \leq \varphi_0) D( p\|   q | \varphi \leq \varphi_0) \nonumber\\&\;\;\;+ \epsilon D(  \varphi p \| \varphi q)\\
	&\leq \varphi_0  D( p\| q ) + \epsilon D(  \varphi p \| \varphi q)\\
	&\leq  \frac{\varphi_{0}}{1-\epsilon}   D( p\|  q ).
\end{align}
Thus we see that $\varphi_{\max}$ (the true maximum) can be replaced by $\varphi_{\text{effmax}}=\frac{\varphi_{0}}{1-\epsilon}$ (an `effective' maximum), for each choice of $\epsilon$ and corresponding $\varphi_0$. Since $\epsilon$ determines the choice of $\varphi_0$, we can in principle optimise the choice of $\epsilon$ in order to get the best possible bound. 

Note that this proof works also for the finite case, thus can be useful when $A$, albeit bounded, takes unconveniently large values with low probability.

\section{Thermodynamic Uncertainty Relations}

Thermodynamics Uncertainty Relations are concerned with upper bounds on the ratio
$|\langle f \rangle|/\sqrt{\langle f^2 \rangle}$ (or, equivalently, $|\langle f \rangle|/\text{StdDev} f$ where StdDev is the standard deviation), for a time-antisymmetric observable $f$.

For a constant-rate or time-symmetric-rate Markov chain, we find \cite{falasco2020unifying} that the time-antisymmetric observable with highest such ratio is 
\begin{align}
	f^*(\omega)=\frac{1}{2} \frac{p(\omega)-p(\overline{\omega})}{p(\omega)+p(\overline{\omega})}.
\end{align}
Indeed from Cauchy-Schwartz inequality we find, for any time-antisymmetric observable $f$:

\begin{align}
	\langle f \rangle^2 = \langle f f^* \rangle^2 \leq \langle f^2 \rangle^2 \langle (f^*)^2 \rangle
\end{align}
with equality for $f=f^*$. Thus we find 

\begin{align}
	\langle f \rangle^2 = \langle f f^* \rangle^2 \leq \langle f^2 \rangle^2 \langle (f^*)^2 \rangle
\end{align}

Moreover we observe that $\langle f^* \rangle=\langle (f^*)^2 \rangle=\langle f^* \rangle^2/\langle (f^*)^2 \rangle$  and $\langle |f^*| \rangle=d_{TV}(p,\tilde{p})$ and $|f^*|\leq 1/2$.

From \eqref{eq:crsigmahantisym} with $f^*$ we thus find

\begin{align}\label{eq:ourTURstar}
	\Delta \sigma   + D(p_{X_\text{fin}}\|p_{X_\text{ini}})  \geq 4\langle f^* \rangle_p \text{tanh}\left(\frac{\langle f^* \rangle_p}{d_{TV}(p,\tilde{p})} \right).	
\end{align}

Since for any other time-antisymmetric observable we have $|\langle f \rangle_p| / \sqrt{\langle f^2 \rangle_p} \leq \langle f^* \rangle=\langle (f^*)^2 \rangle$, we thus find \eqref{eq:ourTUR}.

Note that when $p_e/p_{\overline{e}} \to \infty$ or $0$ for every transition (infinitesimal-time trajectory) $e$, we find $\langle f^* \rangle \to  \langle |f^*| \rangle = d_{TV}(p,\tilde{p})$, thus \eqref{eq:ourTUR} correctly predicts an infinite entropy production, unlike \eqref{eq:classicTUR} which remains bounded.

\end{document}